\documentclass[11pt,a4paper]{article}

\usepackage{amsmath}
\usepackage{graphicx}
\usepackage{hyperref}
\usepackage[hyphenbreaks]{breakurl}
\usepackage{amsmath}
\usepackage{amssymb}
\usepackage{epsfig}
\usepackage{afterpage}
\usepackage{changepage}
\usepackage{amsmath}
\usepackage[normalem]{ulem}

\usepackage{placeins}
\usepackage{bm}

\newcommand{\bcln}{b \to c l^- \bar{\nu}_{l}}

\def \bt{B \to X_c \tau^- \bar{\nu}_{\tau}}
\def \btq{b \to X \tau^- \bar{\nu}_{\tau}}

\def \btuq{b \to X_u \tau^- \bar{\nu}_{\tau}}

\def \bl{B \to X_c \ell^- \bar{\nu}_{\ell}}

\def \beq{\begin{equation}}
\def \eeq{\end{equation}}
\def \bea{\begin{eqnarray}}
\def \eea{\end{eqnarray}}
\def \ber{\begin{eqnarray*}}
\def \eer{\end{eqnarray*}}
\def \bwt{\begin{widetext}}
\def \ewt{\end{widetext}}
\def \nn{\nonumber}

\def \RD{R({D^{(*)}})}

\def \RDr{R_D^{Ratio}}

\def \RDrstar{R_{D^\ast}^{Ratio}}

\def \RXc{\frac{\Gamma(B \to X_c\tau \bar{\nu}_\tau)}{\Gamma(B \to X_c \ell \bar{\nu}_\ell)}}

\def \B{\frac{d\Gamma(B \to X_c\tau \bar{\nu}_\tau)/dq^2}{d\Gamma(B \to X_c \ell \bar{\nu}_\ell)/dq^2}}

\newcommand{\bctaunutau}{b \to c \tau^- {\bar\nu}_\tau}

\def\bra#1{\left\langle #1\right|}
\def\ket#1{\left| #1\right\rangle}

\def \({\left(}
\def \){\right)}
\def \[{\left[}
\def \]{\right]}
\def \l|{\left|}
\def \r|{\right|}

\def \nn{\nonumber}
\def \nl{\nn \\}

\begin{document}

{\flushright
UMISS-HEP-2018-01 \\
}

\begin{center}
\bigskip
{ \Large \bf \boldmath New physics in inclusive $B \to X_c\ell \bar{\nu}$ decay in light of $\RD$ measurements
} \\
\bigskip
\bigskip
{\large

 Saeed Kamali$^{\dagger\zeta}$ $^{}$\footnote{skamali@go.olemiss.edu},  Ahmed Rashed$^{\dagger\ddagger}$ $^{}$\footnote{amrashed@go.olemiss.edu} and   Alakabha Datta$^{\dagger\zeta}$ $^{}$\footnote{datta@phy.olemiss.edu}
      \\
}
\end{center}

\begin{flushleft}

~~~~~~~~~~~ {$^\dagger$\it Department of Physics and
  Astronomy, University of Mississippi, \centerline{108 Lewis Hall, University, MS 38677-1848, USA} } \\
~~~~~~~~~~~ {$^\zeta$\it Department of Physics and Astronomy, University of California, \centerline{Irvine, CA 92697, USA}}\\
~~~~~~~~~~~~ {$^\ddagger$\it Department of Physics, Faculty of Science, Ain Shams University, \centerline{ Cairo, 11566, Egypt } } \\

\end{flushleft}

\begin{center}

\bigskip (\today)
\vskip0.5cm {\Large Abstract\\} \vskip3truemm
\parbox[t]{\textwidth}{In this work we study the effects of new-physics (NP) operators with different Lorentz structures on the inclusive  $B \to X_c\tau \bar{\nu}$ decay and make  predictions for
 the ratio of total decay rates $R(X_c)=\RXc$ with $\ell=e, \mu$, the differential decay rates, $\frac{d\Gamma}{dq^2}$ and $\frac{d\Gamma}{dE_\tau}$, forward-backward asymmetry $A_{FB}$ and the ratio of the differential decay rates $B(q^2)=\B$.
   In addition, we introduce 
some leptoquark models as explicit models of the NP operators and study their effects on the inclusive decay. We consider $\mathcal{O}(\alpha_s)$ radiative and $1/m_b$ nonperturbative corrections to the Standard Model (SM) decay rate  
and ignore their small effects in the NP contributions.}
\end{center}

\thispagestyle{empty}
\newpage
\setcounter{page}{1}
\baselineskip=14pt

\section{Introduction}
\label{sec:introduction}
The $b\to c \tau \nu$ transitions have attracted a lot of attention recently as there is an excess of $\tau$ production compared to SM predictions. 
The observables where the discrepancies are measured, are  the ratios of branching fractions of the semileptonic decays $\bar{B} \to D^{(*)}$ defined by $R(D^{(*)})=\mathcal{B}(\bar{B} \to D^{(*)} \tau^- \bar{\nu}_{\tau})/\mathcal{B}(\bar{B} \to D^{(*)} \ell^- \bar{\nu}_{\ell})$, where $\ell=e,\mu$. These ratios have been measured by the BABAR \cite{Lees:2013uzd, Lees:2012xj}, Belle \cite{Huschle:2015rga, Abdesselam:2016cgx, Sato:2016svk, Hirose:2016wfn}, and LHCb \cite{Aaij:2015yra} collaborations and their values are found to be considerably larger than their SM predictions.
These ratios of branching fractions have certain advantages over the absolute branching fraction measurement
of $ B \to D^{(\ast)} \tau  \nu_\tau$ decays, as they are  relatively less sensitive to form factor variations and several systematic uncertainties,
such as those on the experimental efficiency, as well as the dependence on the value of $|V_{cb}|$, cancel in the ratios.
We take the SM predictions for these ratios to be (for $\ell=e$)
\begin{align}
R(D)_{SM}=0.298 \pm 0.003,  \nonumber\\
R(D^*)_{SM}=0.255 \pm 0.004.
\label{SMRD}
\end{align}
There are lattice QCD predictions for the ratio $R(D)_{SM}$ in the Standard Model \cite{Bailey:2012jg,Lattice:2015rga,Na:2015kha} that are in good agreement with one another,
\bea
R(D)_{SM} &=& 0.299 \pm 0.011 \quad \quad [\mathrm{FNAL/MILC}], \\
R(D)_{SM} &=& 0.300\pm 0.008 \quad \quad [\mathrm{HPQCD}].
\eea
Combining experimental and lattice results, the authors of Ref.~\cite{Bigi:2016mdz} obtained  $R(D)_{SM}$ in Eq.(\ref{SMRD}) (with $\ell=\mu$)
which is also in good agreement with the phenomenological prediction in 
Ref.~\cite{Sakaki:2013bfa}.

{A calculation of ${R}(D^\ast)_{SM}$ is not yet available from lattice QCD and hence  one can use a phenomenological prediction using form factors extracted from $B\to D^*\ell\bar{\nu}$ experimental data \cite{ Sakaki:2013bfa, Fajfer:2012vx}.
Recently there have been new analyses of SM predictions of $R(D^{*})$ \cite{Bigi:2017jbd, Bernlochner:2017jka, Jaiswal:2017rve}. Here we use the results of \cite{Jaiswal:2017rve} (where they do a combined analysis of the experimental data, lattice QCD and light cone sum rule results) to produce the SM prediction in (\ref{SMRD}).
%



%

The averages of $R(D)$ and $R(D^*)$ measurements evaluated by 
the Heavy-Flavor Averaging Group are \cite{FPCP 2017}
\begin{align}
\label{RDexp}
R(D)_{exp}&=0.407 \pm 0.039 \pm 0.024,   \\
\label{RDsexp}
R(D^*)_{exp}&=0.304 \pm 0.013 \pm 0.007. 
\end{align}
The combined analysis of $R(D)$ and $R(D^*)$, taking into account measurement correlations, finds that the deviation is at the level of $4.1 \sigma$ from the SM prediction \cite{FPCP 2017}.
These measurements could be a signal of a new physics beyond SM where  the coupling of new physics to the lepton generation is not universal.
It is logical to probe this new physics in other decay modes which are connected to the $\RD$ anomalies via the same
 $b\to c \tau \nu$ quark level transitions.  An example of such a connected decay is the inclusive decay $\bt$.

In this work we study the effect of new-physics (NP) operators with different Lorentz structures on the inclusive  decay $\bt$.  In a model independent approach we consider the most general dimension-6 new-physics operators that contribute to
the $\bctaunutau$  decay. 
In our calculations we construct the ratios of the experimental results  (\ref{RDexp}) and (\ref{RDsexp}) to the phenomenological SM predictions:
\begin{align}
R(D)^{Ratio}=\frac{R(D)_{exp}}{R(D)_{SM}}=1.36 \pm 0.15, \nonumber\\
R(D^*)^{Ratio}=\frac{R(D^*)_{exp}}{R(D^*)_{SM}}=1.19 \pm 0.06.
\label{ratios}
\end{align} 
We use these values in Eq.~(\ref{ratios}) to find the allowed parameter space of the NP models as done in Ref. \cite{Datta:2017aue, Lam}.
Taking one operator at a time we fix the size of the operators by fitting to the measurements in   Eq.~(\ref{ratios}) \cite{datta1,datta2,datta3} and then  we make predictions for several observables in the inclusive decay. We also consider specific models of new physics where more than one operator is present.
 In recent times, the inclusive decay has been discussed in the SM and with NP by several authors \cite{Ligeti:2014kia, Freytsis:2015qca, Celis:2016azn, Mannel:2017jfk, Grossman:1994ax}. A study of new-physics contributions 
to resolve the tension between the inclusive and exclusive determination of the Cabibbo-Kobayashi-Maskawa (CKM) element $V_{cb}$ in $\bcln$ decays with $\ell= e, \mu$  has been discussed in \cite{Colangelo:2016ymy, Jung:2018lfu}. 

In this paper we assume NP only in $\bctaunutau$ decay and we improve upon the previous calculations in the following way:
\begin{itemize}
\item{We add NP to several differential observables including perturbative $\mathcal{O}(\alpha_s)$ and $1/m_b^2$ power corrections to the SM contribution. These corrections to the forward-backward asymmetry $A_{FB}$ and the ratio of differential rates $B$, have not been worked out previously. Adding $\mathcal{O}(\alpha_s)$ correction to $A_{FB}$ is less trivial than other observables since one has to consider the three-body and four-body decays separately.}  

\item{We consider several leptoquark models where more than one NP coupling is present and study their effects on the inclusive decay $\bt$.}  
\end{itemize}

The theoretical prediction of the inclusive decay rate is rather precise in the SM. The differential decay rate can be expanded systematically both in terms of perturbative and nonperturbative QCD corrections.
Perturbative QCD corrections $\mathcal{O}(\alpha_s)$ to the differential decay rate were calculated in \cite{Czarnecki:1994bn, Jezabek:1996db, Trott:2004xc, AquilaGambino}. For our purpose the calculations in \cite{AquilaGambino} are more useful, where the corrections to the five hadronic structure functions are given and the formulas for the virtual and real gluon corrections are given separately. This allows us to calculate the correction to the phenomenological aspects of the inclusive B decay such as $q^2$ and $E_\tau$ distributions, the ratio of the differential decay rates $B(q^2)=\B$ and more specifically the forward-backward asymmetry, $A_{FB}$.

 Nonperturbative correction to the inclusive semileptonic decay, which is an expansion in $\Lambda_{QCD}/m_b$, is calculated in the context of operator product expansion (OPE) and heavy quark effective theory (HQET) \cite{Balk:1993sz, Koyrakh:1993pq, Falk:1994gw, Blok:1993va}, and \cite{Ligeti:2014kia}.
  Here $m_b$ is the heavy quark mass (bottom quark) and $\Lambda_{QCD}$ is the nonperturbative scale parameter of the strong interactions. In the limit $m_b \to \infty$ we recover the free quark decay and the $\Lambda_{QCD}/m_b$ correction vanishes. The leading order nonperturbative correction is of order $\Lambda_{QCD}^2/m_b^2$ and is parametrized by two hadronic matrix elements, $\lambda_1$ and $\lambda_2$, which are related to the kinetic energy and spin interaction of the $b$ quark in the $B$ meson.
 
Higher order $\mathcal{O}(\alpha_s^2)$ corrections to the total rate are known in the SM, but it turns out that even at order $\mathcal{O}(\alpha_s)$ the radiative corrections to $\bt$ and $\bl$ are correlated and cancel out largely in the ratio of the decay rates 
$R(X_c)= {\mathcal{B}[\bt] \over \mathcal{B}[\bl]}$ \cite{Biswas:2009rb}. So we only consider the order $\mathcal{O}(\alpha_s)$ correction in the ratios of the total/differential decay rates as well as in the definition of the forward-backward asymmetry. The second order QCD corrections to semileptonic $b \rightarrow c$ inclusive transitions, not considered here, can be important for the rates and the absolute differential rates \cite{ Biswas:2009rb,Luke:1994yc} and so the ratios should be considered cleaner  probes of new physics.


Since we consider NP as a correction to the SM, we do not include radiative QCD corrections to the NP part as their contribution should be relatively small. The  effect of nonperturbative $ \mathcal{O}(1/m_b)$ corrections to NP contributions will be considered in a future work.

The paper is organized as follows: The effective Hamiltonian of the NP interactions and helicity amplitudes of the inclusive B decay are presented in Sec.~\ref{sec:formalism}. In Sec.~\ref{corrections}, the power correction and the radiative correction of order $\mathcal{O}(\alpha_s)$ are discussed. The model-independent phenomenological analysis of individual new-physics couplings is considered in Sec.~\ref{sec:modelindependent}, while leptoquark models are considered in Sec.~\ref{sec:models},  and finally we conclude in Sec.~\ref{sec:conclusion}.

\section{Formalism}
\label{sec:formalism}

\subsection{Effective Hamiltonian}

The effective Hamiltonian of the NP operators for the quark-level transition $b\to c\tau^-\bar{\nu}_\tau$  can be written in the form \cite{Chen:2005gr,Bhattacharya:2011qm}
\bea
\label{eq1:Lag}
 {\cal{H}}_{eff} &=&  \frac{G_F V_{cb}}{\sqrt{2}}\Big\{
\Big[\bar{c} \gamma_\mu (1-\gamma_5) b  + g_L \bar{c} \gamma_\mu (1-\gamma_5)  b + g_R \bar{c} \gamma_\mu (1+\gamma_5) b\Big] \bar{\tau} \gamma^\mu(1-\gamma_5) \nu_{\tau} \nl && +  \Big[g_S\bar{c}  b   + g_P \bar{c} \gamma_5 b\Big] \bar{\tau} (1-\gamma_5)\nu_{\tau} + \Big[g_T\bar{c}\sigma^{\mu \nu}(1-\gamma_5)b\Big]\bar{\tau}\sigma_{\mu \nu}(1-\gamma_5)\nu_{\tau} + H.c. \Big\}, \nonumber \\ \label{eq:Heff}
\eea
where $G_F$ is the Fermi constant, $V_{cb}$ is the CKM matrix element, and we use $\sigma_{\mu \nu} = i[\gamma_\mu, \gamma_\nu]/2$. When $g_L = g_R = g_S = g_P = g_T = 0$, the above equation produces the SM effective Hamiltonian.
In this paper, we consider only the active neutrinos which are left chiral. In the presence of new physics, in general, the $\tau$ lepton can be associated with any neutrino flavor. To allow for lepton universality violation we assume NP to dominantly affect the third
generation leptons \cite{datta4, datta5}.
The coefficients of the various operators in the effective Hamiltonian are taken at the $m_b$ energy scale. In general for models defined at a large scale $\Lambda$, one has to run down the Wilson's coefficients to the  $m_b$ energy scale. 
As discussed in Refs.~\cite{Feruglio:2016gvd, Feruglio:2017rjo} the RGE running
 will in general lead to the generation of operators that will lead to other decays and constraints from those decays have to be carefully considered.


\subsection{Decay process}

In this section we present the calculations of the inclusive B decay at the free quark level with new-physics contributions.
The process under consideration is
$$b(p_{b})\rightarrow\tau^{-}(p_{\tau})+\bar{\nu}_{\tau}(p_{\bar{\nu}_\tau})+c(p_{c}).$$
The differential decay rate is
\bea
d\Gamma &=&\frac{1}{2m_b}\frac{G_F^2 |V_{cb}|^2}{4}\sum_{\lambda_c}\sum_{\lambda_\tau}|\mathcal{M}_{\lambda_c}^{\lambda_\tau}|^2 d\Phi_3, \label{eq:rate}
\eea
where $d\Phi_3$ is the three-body phase space which can be written as
\bea
d\Phi_3=\frac{\sqrt{Q_+ Q_-}}{256\pi^3 m_b^2}(1-\frac{m_\tau ^2}{q^2})dq^2 dcos(\theta_\tau),
\eea
with
\bea
q     &=& p_{b}-p_{c}, \\
Q_\pm &=& (m_{b} \pm m_{c})^2 - q^2\,.
\eea
The angle $\theta_\tau$ is defined as the angle between the momenta of the $\tau$ lepton and the $b$ quark in the dilepton rest frame.

The helicity amplitude $\mathcal{M}_{\lambda_{c}}^{\lambda_\tau} $ is written as \cite{Tanaka:2012nw}
\bea
\mathcal{M}^{\lambda_\tau}_{\lambda_{c}}&=&H^{SP}_{\lambda_{c},\lambda=0}L^{\lambda_\tau}+\sum_{\lambda}\eta_{\lambda}H^{VA}_{\lambda_{c},\lambda}L^{\lambda_\tau}_{\lambda}+\sum_{\lambda,\lambda^{\prime}} \eta_\lambda \eta_{\lambda^{\prime}} H^{(T){\lambda_{b}}}_{\lambda_{c},\lambda ,\lambda^{\prime}}L^{\lambda_\tau}_{\lambda,\lambda^\prime}.
\eea
Here, ($\lambda$, $\lambda^\prime$) indicate the helicity of the virtual vector boson, and  $\lambda_{c}$ and $\lambda_\tau$ are the
helicities of the  $c$ quark and $\tau$ lepton, respectively, and $\eta_\lambda=1$ for $\lambda=t$ and $\eta_\lambda=-1$ for $\lambda=0,\pm 1$.

The explicit expressions for the hadronic $(H_{\lambda_{c}})$ and leptonic $(L^{\lambda_\tau})$ helicity amplitudes are presented in Appendix \ref{appendix:Helicity}.

\section{QCD correction to differential decay rates and forward-backward asymmetry}
\label{corrections}
From the twofold decay distribution (\ref{eq:rate}), one may obtain expressions for various observables at the free quark level. These expressions in terms of hadronic helicity amplitudes are presented in Appendix \ref{appendix:distributions}.\\
%
%
Here we shortly discuss the inclusion of QCD corrections to the differential rates. In \cite{AquilaGambino}, the hadronic tensor of the transition $b\rightarrow c$ is parametrized in terms of five hadronic structure functions.  The QCD corrections to these structure functions are calculated to $\mathcal{O}(\alpha_s)$ and generic BLM $(\alpha_s^n \beta_0^{n-1})$ order, and numerical results are given in the massless lepton case.
This correction consists of two parts: loop correction, which is the virtual part and has the same kinematics as the three-body decay, and the real gluon emission (four-body decay) which has an infrared divergence that cancels out with the divergence in the loop contribution.\\ 
Here, using the results of \cite{AquilaGambino}, we add the $\mathcal{O}(\alpha_s)$ correction to the differential decay rates and forward-backward asymmetry in the case where the final lepton is massive.  
To add the $\mathcal{O}(\alpha_s)$ correction, one should find the appropriate integration intervals for the three-body (four-body) decay in the case of loop correction (real gluon emission). Since the correction to the triple differential distribution for $b\rightarrow cl \bar{\nu}_l$ is given as a function of the lepton energy, it is more convenient to introduce the definitions of the forward and backward terms in the forward-backward asymmetry $(A_{FB})$ [Eq.~(\ref{AFB})] in terms of the lepton energy rather than the $\tau$ scattering  angle $\theta_\tau$. Therefore, the integration is done over the lepton energy rather than the angle $\theta_\tau$.

In Appendix \ref{appendix:kinematics} we find the relation between the $\tau$ energy $E_\tau$, which is defined in the $b$ quark's rest frame, and the angle $\theta_\tau$ defined in the dilepton's rest frame. A comprehensive study of decay kinematics is given in \cite{kinematics}.\\
For the energy $E_\tau$ in four-body decay we find (see Appendix \ref{appendix:kinematics})
\bea
E_\tau = \frac{1}{4 m_b q^2}\big[(m_b^2+q^2-r^2)(m_{\tau}^2+q^2)-(q^2-m_{\tau}^2)\sqrt{\lambda(m_b^2,q^2,r^2)}cos(\theta_\tau)\big],
\label{Etau-relation}
\eea
where $\lambda(a,b,c)=a^2+b^2+c^2-2ab-2ac-2bc$, and $q^2$ and $r^2$ are the invariant masses of the dilepton and the charm-quark/gluon systems, respectively. For three-body decay $r^2$ reduces to $m_c^2$. From the above equation we can find the bounds on the $\tau$ energy by $cos(\theta_\tau)=\pm1$,
\begin{equation}
E_\tau^{\pm} = \frac{1}{4 m_b q^2}\big[(m_b^2+q^2-r^2)(m_{\tau}^2+q^2)\pm(q^2-m_{\tau}^2)\sqrt{\lambda(m_b^2,q^2,r^2)}\big].
\end{equation}
Using this equation we can easily calculate the forward-backward asymmetry by performing the integration over $E_\tau$ instead of $cos(\theta_\tau)$. We therefore define the forward-backward asymmetry in the case of four-body decay as
\begin{equation}
A_{FB}=\frac{\int(\int_{E_\tau^-}^{E_\tau^0} \frac{d\Gamma}{dq^2 dr^2 dE_\tau}dE_\tau-\int_{E_\tau^0}^{E_\tau^+} \frac{d\Gamma}{dq^2 dr^2 dE_\tau}dE_\tau)dr^2 }{\frac{d\Gamma}{dq^2}},
\label{AFB}
\end{equation}
where $E_\tau^0=\frac{(m_b^2+q^2-r^2)(m_{\tau}^2+q^2)}{4 m_b q^2}$.  
Note that the integration over $r^2$ appears only in the case of the four-body decay.\\

%
%
%
\section{Model-independent analysis of individual new-physics couplings}
\label{sec:modelindependent}
In this section we consider one NP coupling at a time. Constraints on NP parameters (follow the discussion in Ref.~\cite{Datta:2017aue}) are considered from the existing measurements of $R(D)$ and $R(D^*)$ mesons and from the $B_c$ lifetime. The $B_c$ measurement does not have any significant effect on the constraints except for the $g_P$ coupling. (In general, NP couplings are taken to be complex. Nevertheless, in the numerical analysis of $R(X_c)$, Fig. \ref{fig:model-indp-rate} and Tables~\ref{table1:maxmin} and \ref{table2:maxmin}, we take these couplings to be real for simplicity.)
  
We require the NP couplings to reproduce the measurements of $\RDr$ and $\RDrstar$ within the $3 \sigma$ range.
The coupling $g_S(g_P)$ only contributes to $\RDr (\RDrstar)$ while the other couplings contribute to both channels.
The constraint due to $B_c$ has been considered before in \cite{Li:2016vvp,Alonso:2016oyd,Celis:2016azn}. Here we follow the same procedure and use an upper limit of 
$\mathcal{B}(B_c \to \tau^- \bar{\nu}_\tau) \le 30\%$, and we take $f_{B_c}=0.434(15)\:{\rm GeV}$ from lattice QCD \cite{Colquhoun:2015oha}, to impose this constraint on the NP coupling $g_P$.
A stronger constraint on $\mathcal{B}(B_c \to \tau^- \bar{\nu}_\tau) \le 10\%$ can be obtained  with additional theory input\cite{Ack} which will further constrain the $g_P$ coupling but we will stick to the more conservative estimate
$\mathcal{B}(B_c \to \tau^- \bar{\nu}_\tau) \le 30\%$ in this work.
For numerical inputs we use the $1S$ mass scheme for the quark masses (see \cite{Hoang:1998ng, Hoang:1998hm} and \cite{Ligeti:2014kia}). We use the parameters as given in \cite{Ligeti:2014kia}, and they are presented in Table \ref{params}.


\begin{table}[]
\centering
\begin{tabular}{| c | c |}
\hline
Parameter          & Value                          \\
\hline
\hline
$m_{b}$            & $4.71 \pm 0.05~ GeV$            \\
\hline
$\delta m_{bc}$    & $3.40 \pm 0.02~ GeV$              \\
\hline
$\lambda_1$       &   $-0.30~ GeV^2 \pm 25\%$           \\
\hline
$\lambda_2$       &   $0.12~ GeV^2 \pm 25\%$           \\
\hline
$\alpha_s$        &   $0.218^{+0.065}_{-0.040}$ \\
\hline              
\end{tabular}
\caption{Parameters used in numerical results.}
\label{params}
\end{table}

The SM prediction for the ratio of decay rates becomes 
\begin{equation}
\label{btcSM}
R(X_c)_{SM}=\frac{\mathcal{B}(B\rightarrow X_c \tau \bar{\nu})}{\mathcal{B}(B\rightarrow X_c \ell \bar{\nu})}=0.221 \pm 0.005 ,
\end{equation}
which is comparable with the central value of  $R(X_c)_{SM}=0.222 $ given in \cite{Celis:2016azn} and $R(X_c)_{SM}=0.223$  in \cite{Ligeti:2014kia, Freytsis:2015qca} where they add in addition the $\mathcal{O}(\alpha_s^2)$ correction to the total rate.\\
Recently power correction of order $1/m_b^3$ to this decay rate has been calculated \cite{Mannel:2017jfk}. Taking into account this correction will result in a reduction of $\sim 7\%$ in $R(X_c)$ which is a noticeable effect. Nevertheless, in order to be consistent throughout the paper we do not consider this correction for our numerical study and we present all observables calculated up to the same perturbative and nonperturbative order.\\ 

We now consider the effect of NP models on the total inclusive decay rate.  There is an ALEPH measurement \cite{ALEPH}
\begin{equation}
\mathcal{B}(\btq)_{exp}=(2.43 \pm 0.32) \times 10^{-2},
\end{equation}
where $X=X_c+X_u$ are all possible states from $b \to c$ and $b \to u$ transitions.
In some part of our analysis we will use the above measurement as an experimental input. When we do that we will set the ALEPH measurement  to the inclusive rate for  $\mathcal{B}(\bt) $. The ALEPH measurement represents the inclusive decays of a mixture of $b$ hadrons and in the leading order in the heavy quark expansion all $b$ hadrons have the same width. Moreover, we will neglect the small $\btuq$ transition.

Using the world average for the semileptonic branching ratio into the light lepton \cite{AmhisHFAG2},
\begin{equation}
\mathcal{B}(\bl)_{exp}=(10.65 \pm 0.16) \times 10^{-2},
\end{equation} 
we can find for the ratio 
\begin{equation}
\label{Rxc_exp}
R(X_c)_{exp}=0.228 \pm 0.030 .
\end{equation}

In Fig. \ref{fig:model-indp-rate} we plot the effect of new physics on the ratio of total inclusive decay rates $R(X_c)$ (blue lines) by taking the NP couplings to be real. The pink shaded areas show the allowed range of measured $R(X_c)$, within $1\sigma$ using (\ref{Rxc_exp}) and the green shaded areas are constraints (on the couplings) due to the measurements of $R(D)$ and $R(D^*)$ within a $3\sigma$ interval and the branching ratio of $B_c$. As we can see from the figure, for the $g_S$, $g_L$ and $g_T$ couplings, the experimental $(1\sigma)$ bounds on $R(X_c)$ can reduce the allowed parameter space for the NP couplings. This effect is more pronounced for the $g_L$ and $g_T$ couplings where  a significant part of the allowed coupling values are excluded by $R(X_c)$. The allowed values of the couplings are given in Table \ref{params2}. On the other hand if the ALEPH result is not used as an input, large deviations from the SM are possible for the inclusive rate. As an illustration, in Tables \ref{table1:maxmin} and \ref{table2:maxmin} we present maximum and minimum values of $R(D^{(*)})$ in each model by considering the measurements of $R(D^{(*)})$ and the branching ratio of $B_c$ as constraints, and we compare them with the corresponding values when we add the inclusive measurement as another constraint.

\begin{figure}
\begin{center}
\includegraphics[width=5cm]{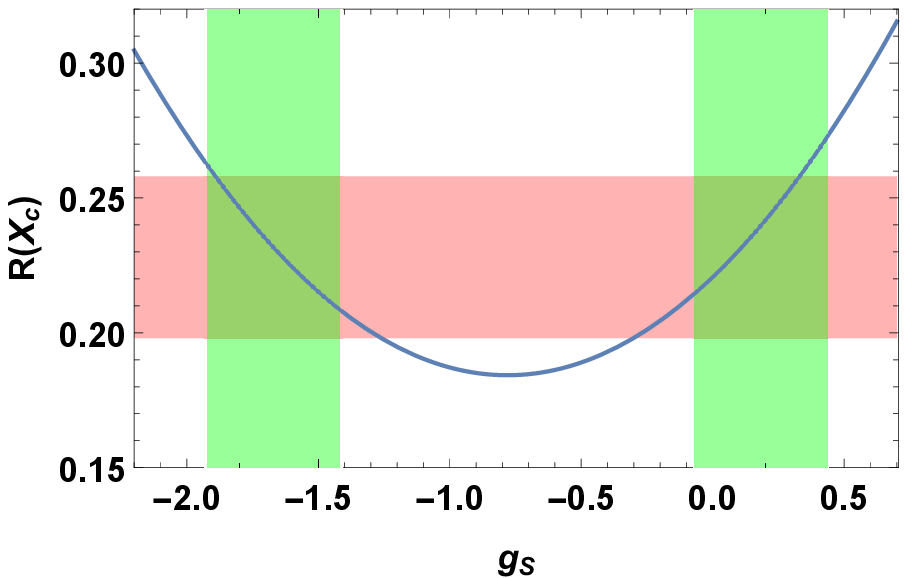}~~~
\includegraphics[width=5cm]{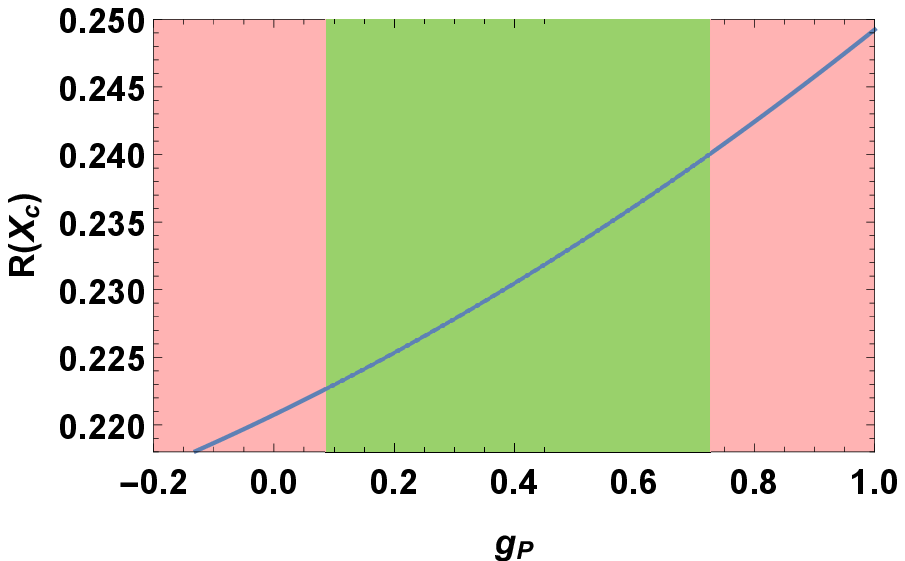}\\[0.3cm]
\end{center}
\begin{adjustwidth}{-0.5cm}{-0.5cm}
\includegraphics[width=5cm]{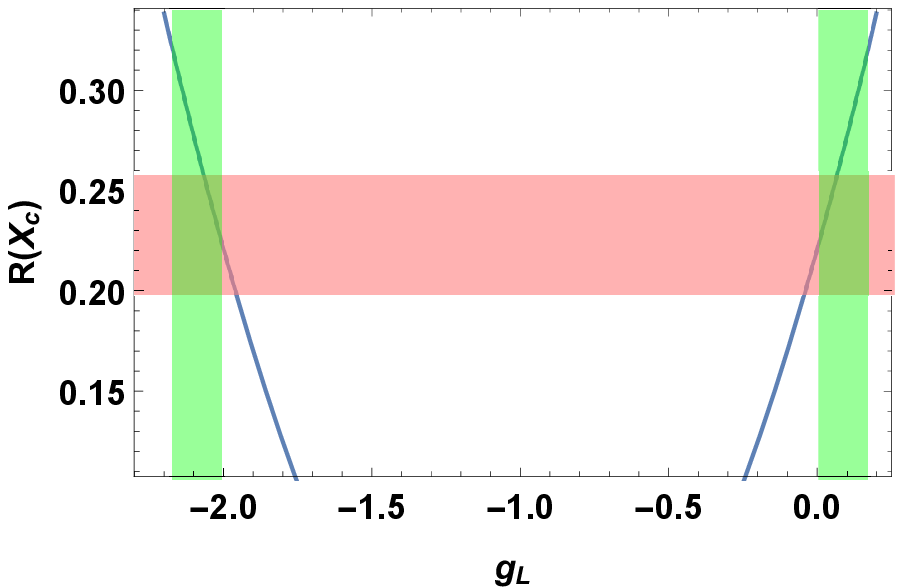}~~~
\includegraphics[width=5cm]{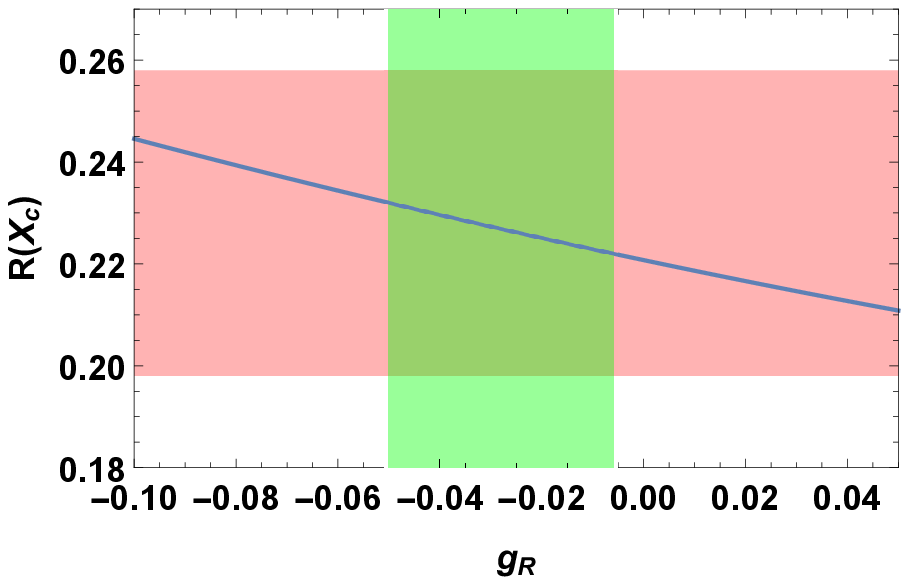}~~~
\includegraphics[width=5cm]{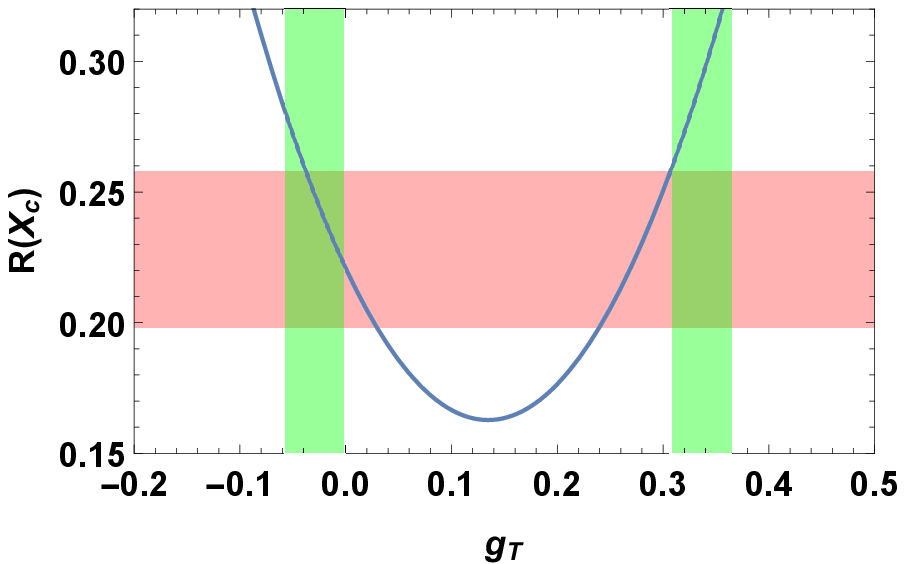}~~~
\end{adjustwidth}
\caption{The effect of real NP couplings on the ratio of total decay rates $R(X_c)$ (blue lines). The pink shaded areas are the allowed regions within $1\sigma$ of the central value for  $R(X_c)_{exp}$ and the green shaded areas are constraints on the couplings due to measurements of $R(D)$ and $R(D^*)$ and the branching ratio of $B_c$.}
\label{fig:model-indp-rate}
\end{figure}

\begin{table}[]
\centering
\begin{tabular}{| c | c |}
\hline
Coupling          & Allowed value                          \\
\hline
\hline
$g_{S}$    & $ (-1.89 , -1.42)$ and $(-0.07 , 0.33)$        \\
\hline
$g_{P}$    & $(0.09 , 0.73)$              \\
\hline
$g_L$       &   $(-2.07 , -2.01)$ and $(0.01 , 0.07)$           \\
\hline
$g_R$       &   $(-0.05 , -0.01)$           \\
\hline
$g_T$        &   $(-0.04 , 0)$  \\
\hline              
\end{tabular}
\caption{Allowed values of the coupling constants taken from Fig. \ref{fig:model-indp-rate}.}
\label{params2}
\end{table}

\begin{table}[]
\centering
\resizebox{\textwidth}{!}{
\begin{tabular}{| l | c | c |}
\hline 
& \multicolumn{1}{c |}{\begin{tabular}[c]{ c }Max/Min Values\\ Without (With) Inclusive Constraint\\ $g_S$ or $g_P$ \end{tabular}}
& \multicolumn{1}{c |}{\begin{tabular}[c]{ c }Max/Min Values\\ Without (With) Inclusive Constraint\\ $g_L$ \end{tabular}}\\
\hline
$R(D)^{Ratio}$ 
& \multicolumn{1}{c |}{\begin{tabular}[c]{ c }1.83/0.90 (1.75/0.90)\\ $at$ $g_S=-1.92$ or $0.43$ / $-1.42$ or $-0.07$ ($g_S=-1.89$/$-1.42$ or $-0.07$ ) \end{tabular}}
& \multicolumn{1}{c |}{\begin{tabular}[c]{ c }1.38/1.01 (1.14/1.01)\\ $at$ $g_L=-2.17$ or $0.17$ / $-2$ or $0.005$ ($g_L=-2.07$ or $0.07$ / $-2$ or $0.005$)  \end{tabular}}\\                                                                                                                                                  
\hline
$R(D)$
& \multicolumn{1}{c |}{\begin{tabular}[c]{ c }0.545/0.269 (0.523/0.269)                                                                                                                                                                                        \\ $at$ $g_S=-1.92$ or $0.43$ / -1.42 or -0.07 ($g_S=-1.89$/$-1.42$ or $-0.07$)  \end{tabular}}
 & \multicolumn{1}{c |}{\begin{tabular}[c]{ c }0.410/0.301 (0.340/0.301)\\ $at$ $g_L=-2.17$ or $0.17$ / $-2$ or $0.005$ ($g_L=-2.07$ or $0.07$ / $-2$ or $0.005$) \end{tabular}}\\                                                                                                
\hline
$R(D^*)^{Ratio}$                                                                                                                         & \multicolumn{1}{c |}{\begin{tabular}[c]{ c }1.10/1.01 (1.10/1.01)                                                                                                  \\ $at$ $g_P=0.726/0.087$ ($g_P=0.726/0.087$)  \end{tabular}} 
& \multicolumn{1}{c |}{\begin{tabular}[c]{ c }1.38/1.01 (1.14/1.01) \\ $at$ $g_L=-2.17$ or $0.17$ / $-2$ or $0.005$ ($g_L=-2.07$ or $0.07$ / $-2$ or $0.005$)  \end{tabular}} \\                                                                                                   
\hline
$R(D^*)$                                                                                                                                                               & \multicolumn{1}{c |}{\begin{tabular}[c]{ c }0.281/0.257 (0.281/0.257)                                                                                               \\ $at$ $g_P=0.726/0.087$ ($g_P=0.726/0.087$)  \end{tabular}} 
 & \multicolumn{1}{c |}{\begin{tabular}[c]{ c }0.351/0.257 (0.290/0.257) \\ $at$ $g_L=-2.17$ or $0.17$ / $-2$ or $0.005$ ($g_L=-2.07$ or $0.07$ / $-2$ or $0.005$)  \end{tabular}}\\                                                                                                
\hline
\end{tabular}
}
\caption{Comparing maximum and minimum values of $R(D^{(*)})$ by using measurements of $R(D^{(*)})$ and the branching ratio of $B_c$ without (with) adding the inclusive measurement as a constraint.}
\label{table1:maxmin}
\end{table}

\begin{table}[]
\centering
\resizebox{\textwidth}{!}{
\begin{tabular}{| l | c | c |}
\hline
& \multicolumn{1}{c |}{\begin{tabular}[c]{ c }Max/Min Values\\ Without (With) Inclusive Constraint\\ $g_R$ \end{tabular}}
& \multicolumn{1}{c |}{\begin{tabular}[c]{c}Max/Min Values\\ Without (With) Inclusive Constraint\\ $g_T$ \end{tabular}}\\
\hline
$R(D)^{Ratio}$                                                                                            & \multicolumn{1}{c |}{\begin{tabular}[c]{ c }0.99/0.90 (0.99/0.90)                                                                           \\ $at$ $g_R=-0.006/-0.05$ ($g_R=-0.006/-0.05$) \end{tabular}} 
& \multicolumn{1}{c |}{\begin{tabular}[c]{ c }1.41/0.95 (1.00/0.97) \\ $at$ $g_T=0.365$/$-0.058$ ($g_T=-0.002/-0.038$)  \end{tabular}}\\                                                                              
\hline
$R(D)$
& \multicolumn{1}{c |}{\begin{tabular}[c]{ c }0.295/0.269 (0.295/0.269)                                                                                              \\ $at$ $g_R=-0.006/-0.05$ ($g_R=-0.006/-0.05$)  \end{tabular}}
 & \multicolumn{1}{c |}{\begin{tabular}[c]{ c }0.421/0.283 (0.298/0.288) \\ $at$ $g_T=0.365$/$-0.058$ ($g_T=-0.002/-0.038$) \end{tabular}}\\                                                                           
\hline
$R(D^*)^{Ratio}$                                                                                                                          & \multicolumn{1}{c |}{\begin{tabular}[c]{ c }1.09/1.01 (1.09/1.01)                                                                                                   \\ $at$ $g_R=-0.05/-0.006$ ($g_R=-0.05/-0.006$)  \end{tabular}} 
& \multicolumn{1}{c |}{\begin{tabular}[c]{ c }1.38/1.01 (1.23/1.01) \\ $at$ $g_T=0.365$ or $-0.058$ / $0.309$ or $-0.002$ ($g_T=-0.038/-0.002$)  \end{tabular}}\\                                                                               
\hline
$R(D^*)$                                                                                                                                                                & \multicolumn{1}{c |}{\begin{tabular}[c]{ c }0.278/0.257 (0.278/0.257)                                                                                               \\$at$ $g_R=-0.05/-0.006$ ($g_R=-0.05/-0.006$)  \end{tabular}}
 & \multicolumn{1}{c |}{\begin{tabular}[c]{ c }0.351/0.257 (0.314/0.257) \\ $at$ $g_T=0.365$ or $-0.058$ / $0.309$ or $-0.002$ ($g_T=-0.038/-0.002$) \end{tabular}}\\                                                                      
\hline
\end{tabular}
}
\caption{Comparing maximum and minimum values of $R(D^{(*)})$ by using measurements of $R(D^{(*)})$ and the branching ratio of $B_c$ without (with) adding the inclusive measurement as a constraint.}
\label{table2:maxmin}
\end{table}

Now we consider differential rates and we first  consider effects of perturbative and nonperturbative corrections to the differential rates in the SM. In Fig.~\ref{fig:comparisons} we plot the differential distributions, $\frac{1}{\Gamma_0}\frac{d\Gamma}{dq^2}$ and $\frac{1}{\Gamma_0}\frac{d\Gamma}{dE_\tau}$, the ratio of the differential decay rate $B=\B$, and the forward-backward asymmetry $A_{FB}$ in Eq.~(\ref{AFB}) in the SM at leading and next-to-leading order and with the $1/m_b^2$ correction. We normalize these observables to $\Gamma_0$ where
\begin{equation}
\Gamma_0=\frac{G_F^2 |V_{cb}|^2 m_b^5}{192 \pi^3}.
\end{equation}
As shown, the radiative correction to $B$ and $A_{FB}$ is not as effective as in the case of $d\Gamma/dq^2$ or $d\Gamma/dE_\tau$.
In general, we expect higher order perturbative corrections to affect the $q^2$ and the $E_\tau$ distributions by  larger amounts compared to the  $B$ and the $A_{FB}$ observables which involve ratios of differential quantities.
The $1/m_b^2$ correction has a considerable effect on all observables except the ratio of differential branching ratios, $B$. In this observable the power correction becomes noticeable only close to the end point region. In general however, one should be careful when interpreting the power corrections locally as the OPE breaks down near the end points.  


\begin{figure}
\begin{center}
\includegraphics[width=5cm]{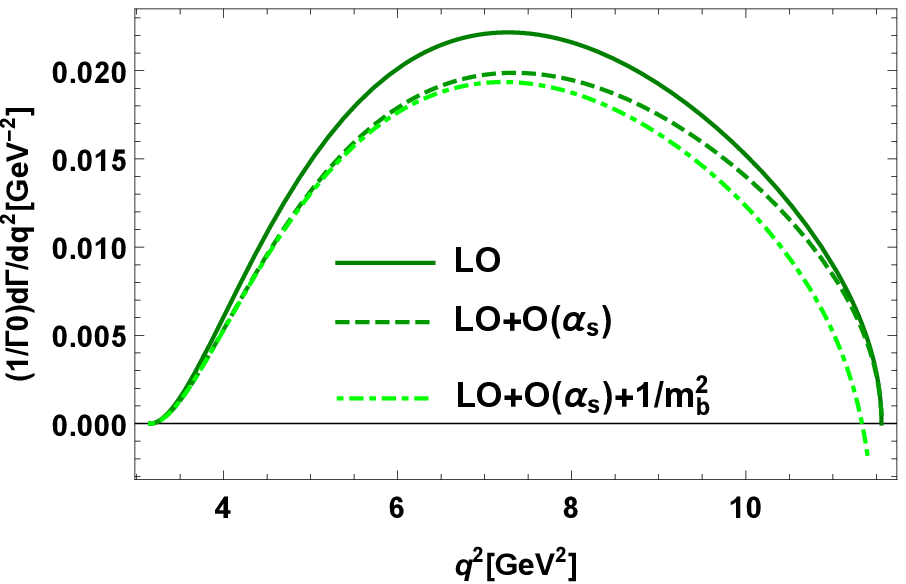}~~~
\includegraphics[width=5cm]{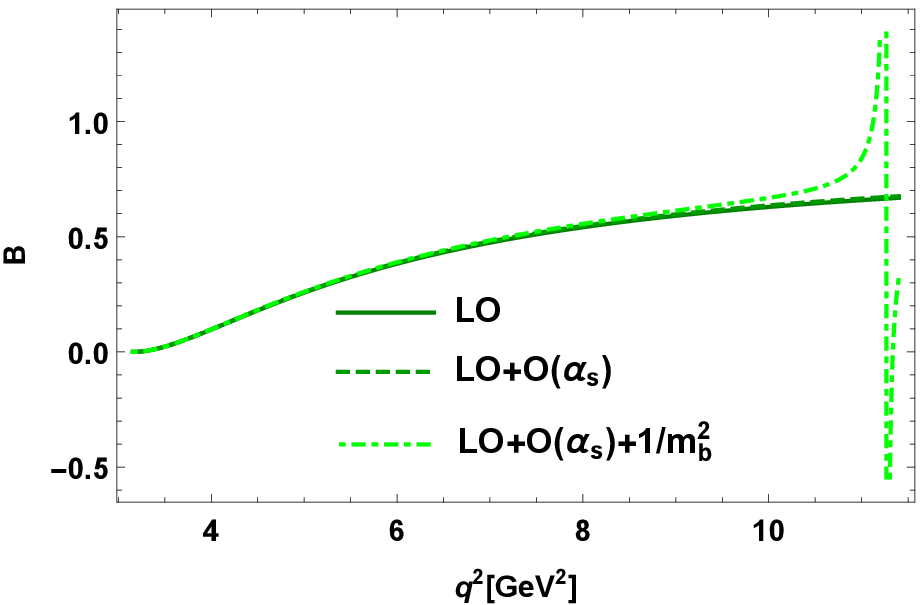}~~~\\
\includegraphics[width=5cm]{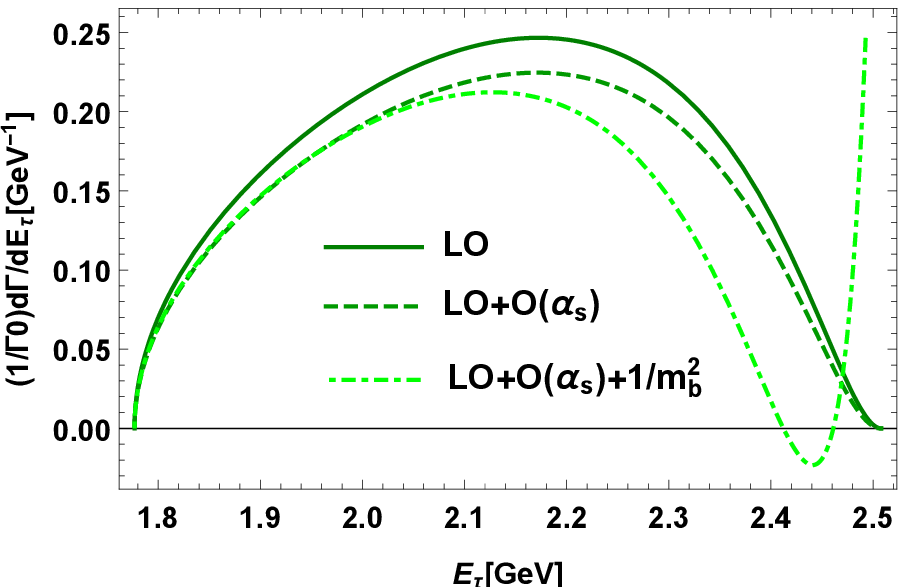}~~~
\includegraphics[width=5cm]{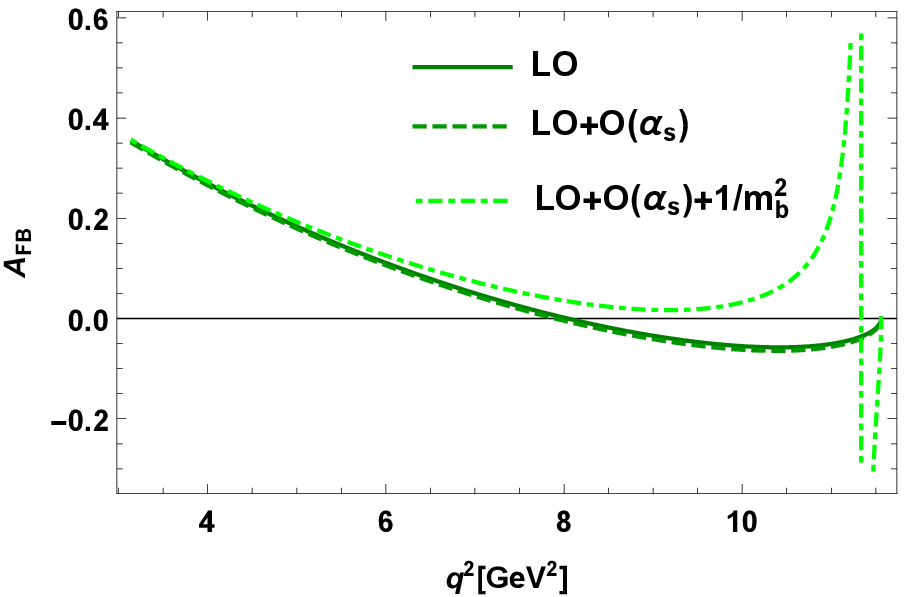}
\end{center}
\caption{The differential decay rates $(1/\Gamma_0)d\Gamma/dq^2$ and $(1/\Gamma_0)d\Gamma/dE_\tau$, the ratio of the differential decay rates $B$, and forward-backward asymmetry $A_{FB}$ at leading (solid line), next-to-leading (dashed line) and next-to-leading order with $1/m_b^2$ correction (dashed-dotted line) for the process $\bt$.}
\label{fig:comparisons}
\end{figure}

In Figures \ref{fig:individualcouplingexamples1} - \ref{fig:individualcouplingexamples2} we present the effects of different NP couplings on the observables $\frac{1}{\Gamma_0}\frac{d\Gamma}{dq^2}$, $\frac{1}{\Gamma_0}\frac{d\Gamma}{dE_\tau}$, $B$, and $A_{FB}$ by considering one coupling at a time. In these plots, the SM uncertainties to the observables are shown as error bars.
To calculate these uncertainties we use the numerical values in Table \ref{params} and propagate the uncertainties for each observable. To account for $\mathcal{O}(\alpha_s^2)$ corrections for each observable, we use the calculations in \cite{Biswas:2009rb} where the $\mathcal{O}(\alpha_s)$ and $\mathcal{O}(\alpha_s^2)$ orders contribute to the total decay rate with the amount of about $10\%$ and $6\%$ of the leading order, respectively. Therefore, we assume the unknown higher order 
contributions in the differential distributions to follow the same ratios. We estimate the errors due to $\mathcal{O}(\alpha_s^2)$ corrections to be $ \pm 70\%$ of the $\mathcal{O}(\alpha_s)$ correction and add this estimate as an uncertainty to the differential decay rates.
For the two observables $B$ and $A_{FB}$, we see that these uncertainties are considerably smaller.\\
Except for the $g_P$ coupling which is tightly constrained by $B_c$, we see that NP models can have considerable effects on these observables in general.  In particular we see that $A_{FB}$ can have zero crossings and take negative values unlike the SM  for some NP couplings. The imaginary parts if present in the NP couplings lead to $CP$ violation in the inclusive $B$ decay. However, the presence of $CP$ violation like rate asymmetries requires the existence of weak phases as well as strong phases. This is a separate point to be discussed in another paper.

%

\begin{figure}
\begin{center}
\includegraphics[width=5cm]{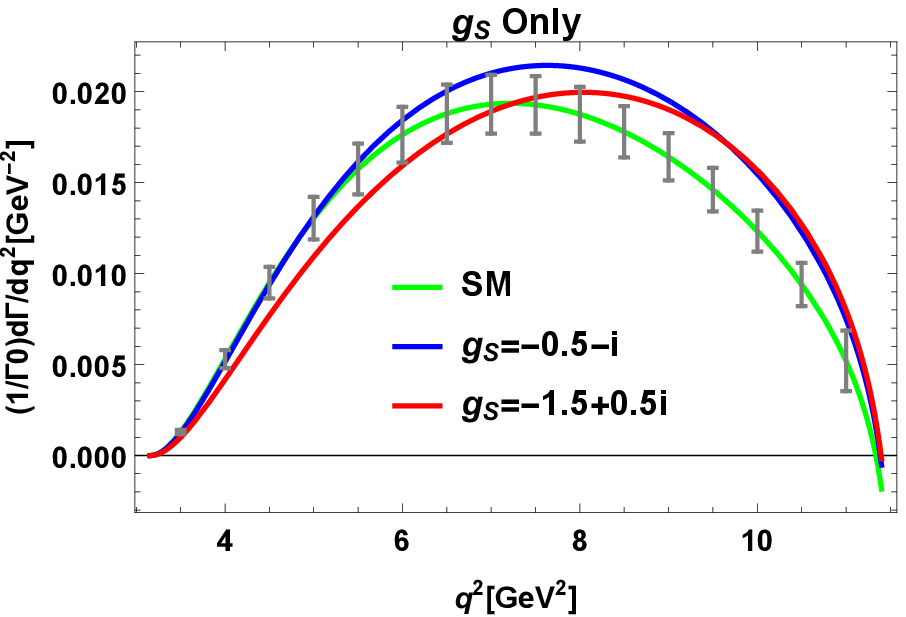}~~~
\includegraphics[width=5cm]{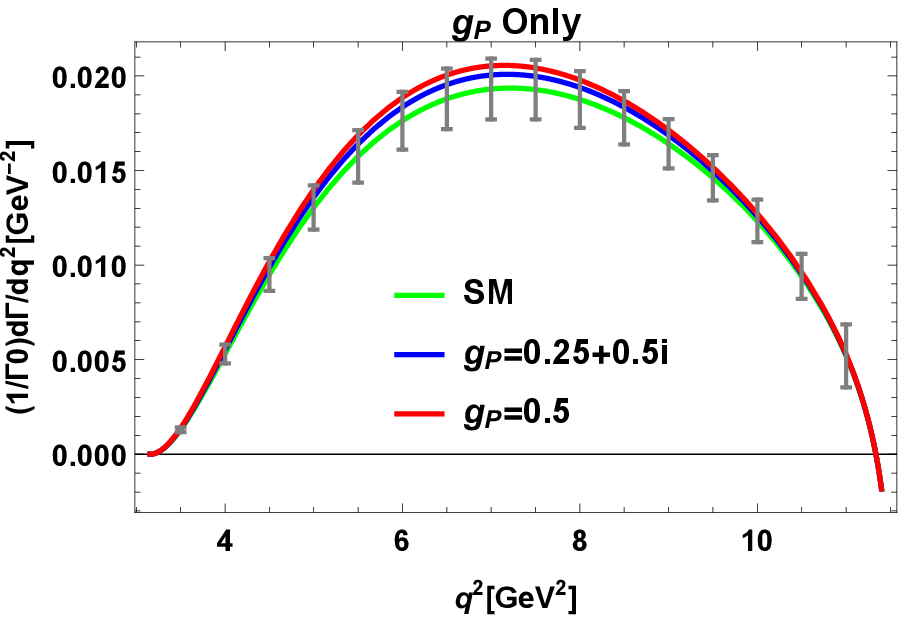}\\[0.3cm]
\end{center}
\begin{adjustwidth}{-0.5cm}{-0.5cm}
\includegraphics[width=5cm]{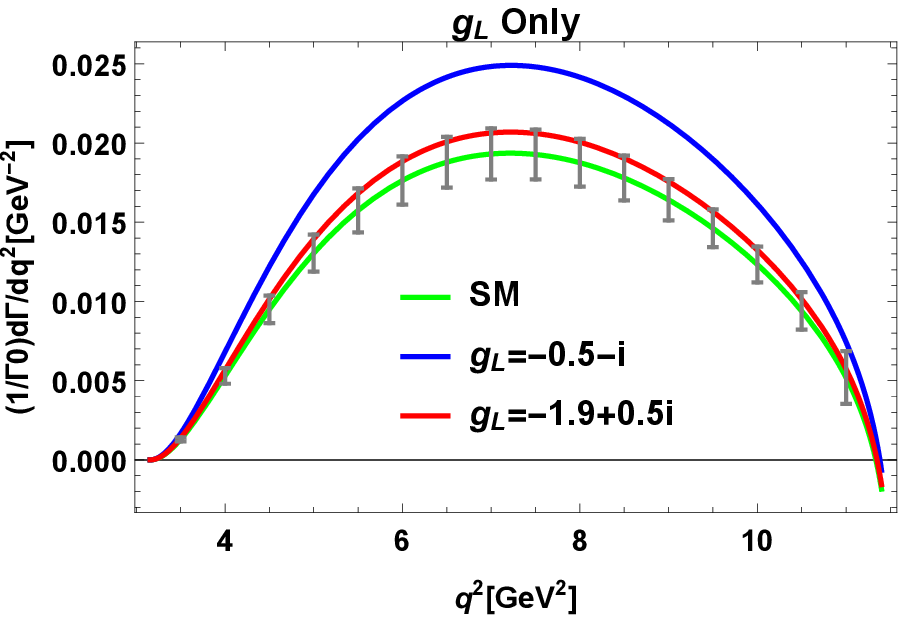}~~~
\includegraphics[width=5cm]{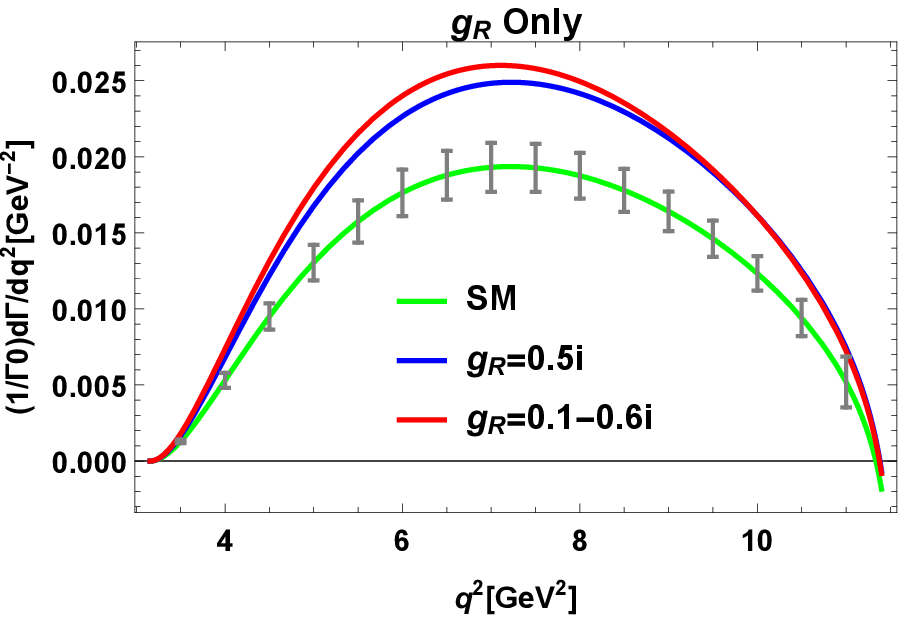}~~~
\includegraphics[width=5cm]{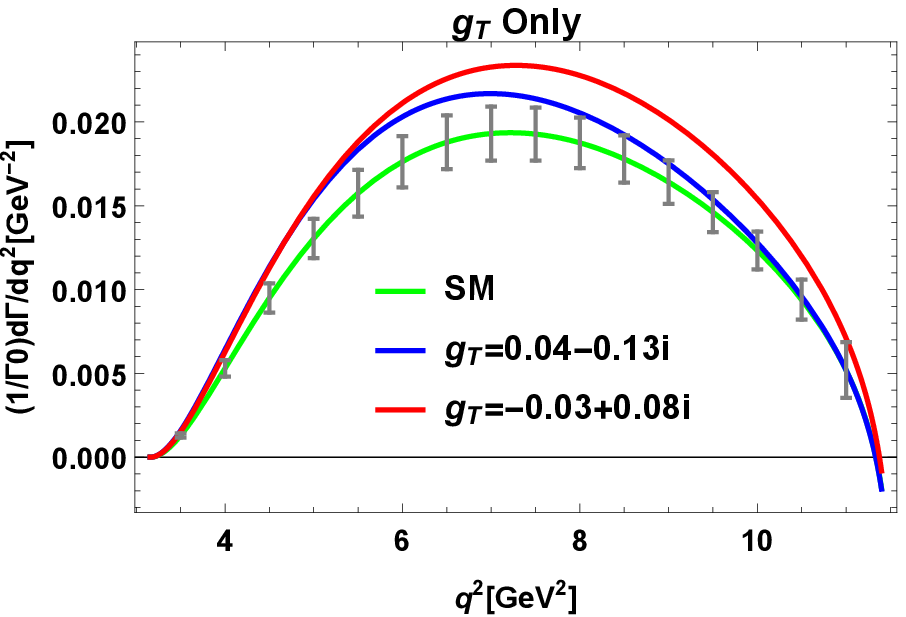}\\[0.3cm]
\end{adjustwidth}
\caption{The effect of individual new-physics couplings on the $\bt$ differential decay rate $(1/\Gamma_0)d\Gamma/dq^2$, including the QCD $\mathcal{O}(\alpha_s)$ and $1/m_b^2$ correction in the SM contribution only. Each plot shows the observable in the Standard Model and for two allowed values of the new-physics couplings.}
\label{fig:individualcouplingexamples1}
\end{figure}

\begin{figure}
\begin{center}
\includegraphics[width=5cm]{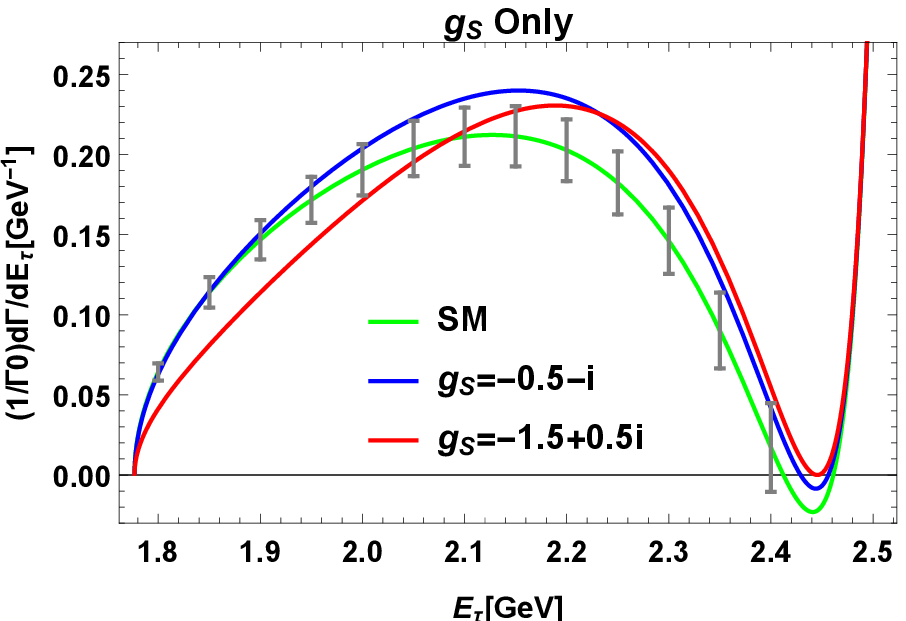}~~~
\includegraphics[width=5cm]{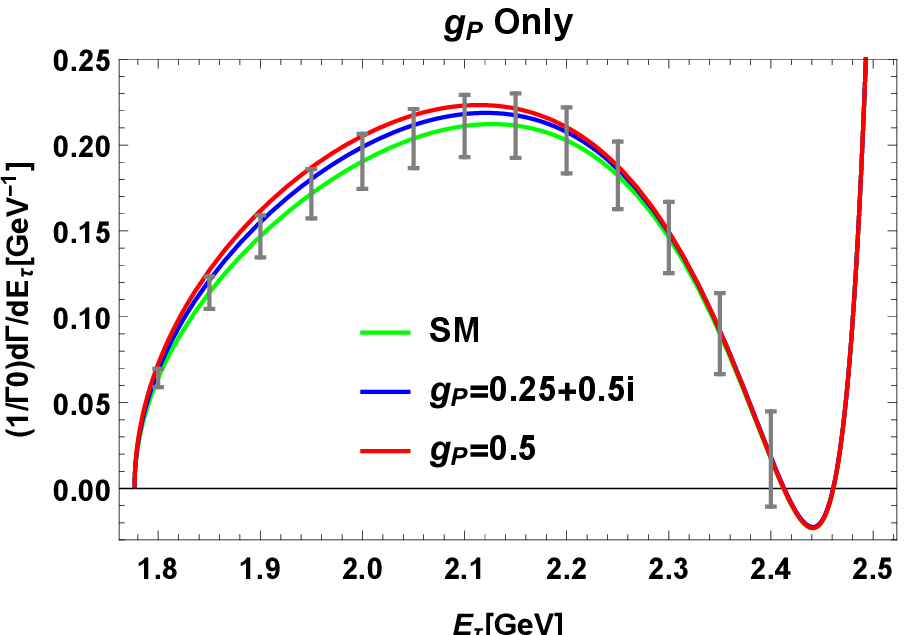}\\[0.3cm]
\end{center}
\begin{adjustwidth}{-0.5cm}{-0.5cm}
\includegraphics[width=5cm]{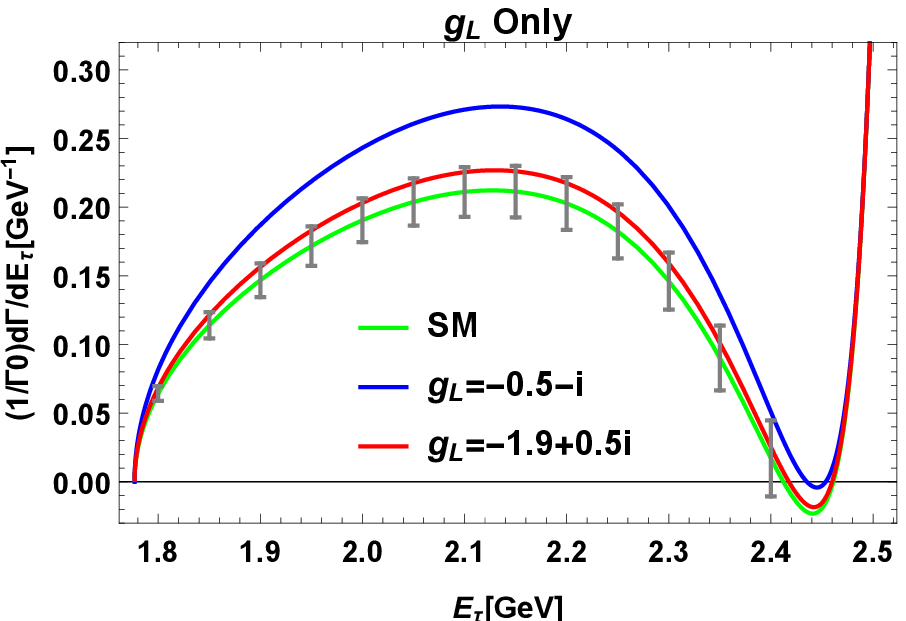}~~~
\includegraphics[width=5cm]{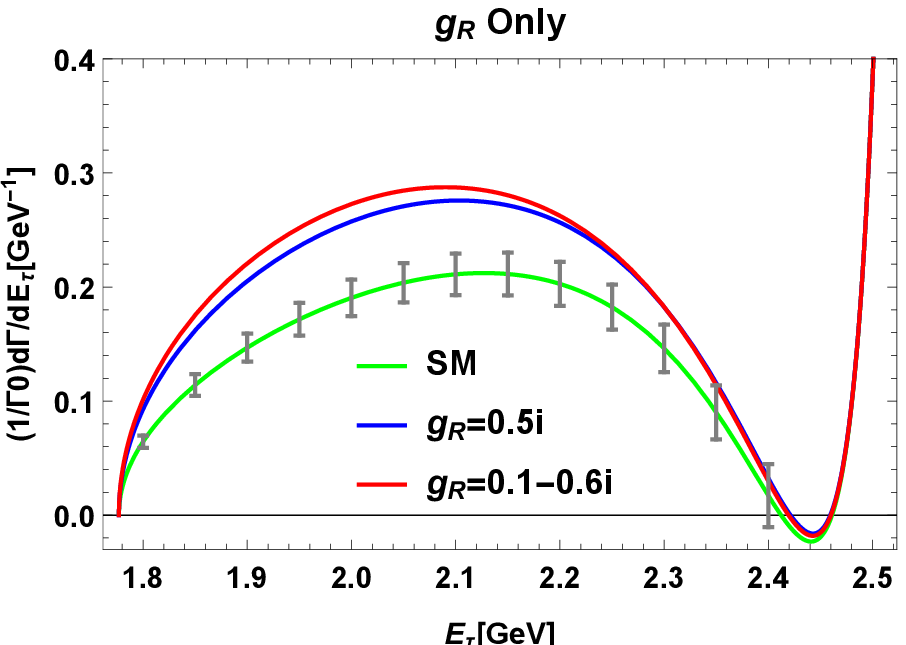}~~~
\includegraphics[width=5cm]{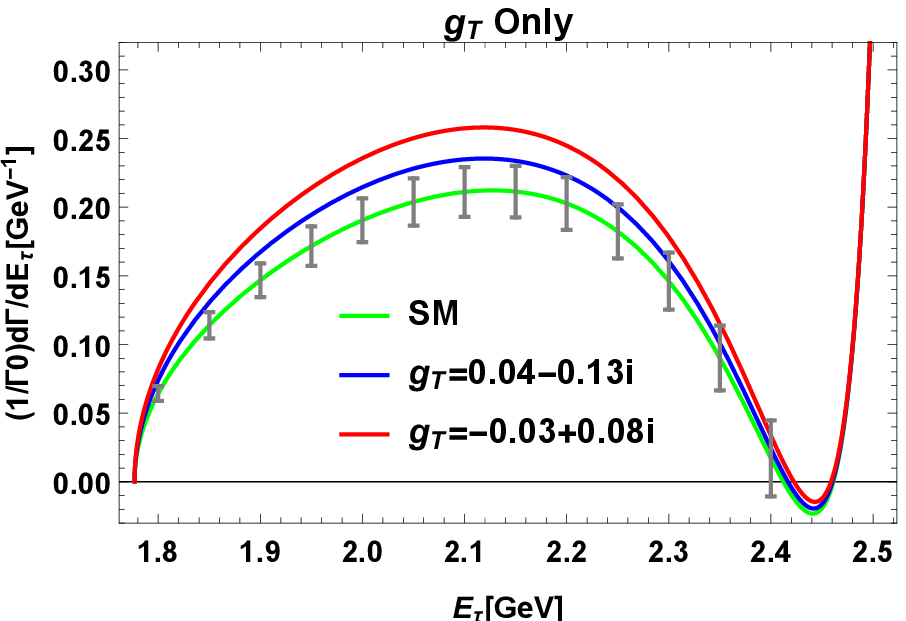}\\[0.3cm]
\end{adjustwidth}
\caption{The effect of individual new-physics couplings on the $\bt$ differential decay rate $(1/\Gamma_0)d\Gamma/dE_\tau$, including the QCD $\mathcal{O}(\alpha_s)$ and $1/m_b^2$ correction in the SM contribution only. Each plot shows the observable in the Standard Model and for two allowed values of the new-physics couplings.}
\label{fig:individualcouplingexamples11}
\end{figure}

\begin{figure}
\begin{center}
\includegraphics[width=5cm]{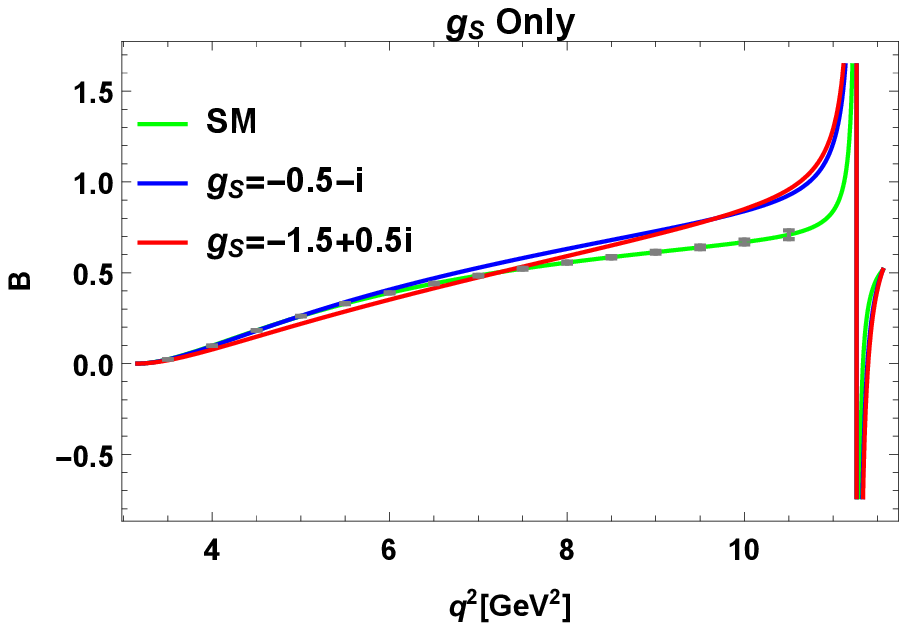}~~~
\includegraphics[width=5cm]{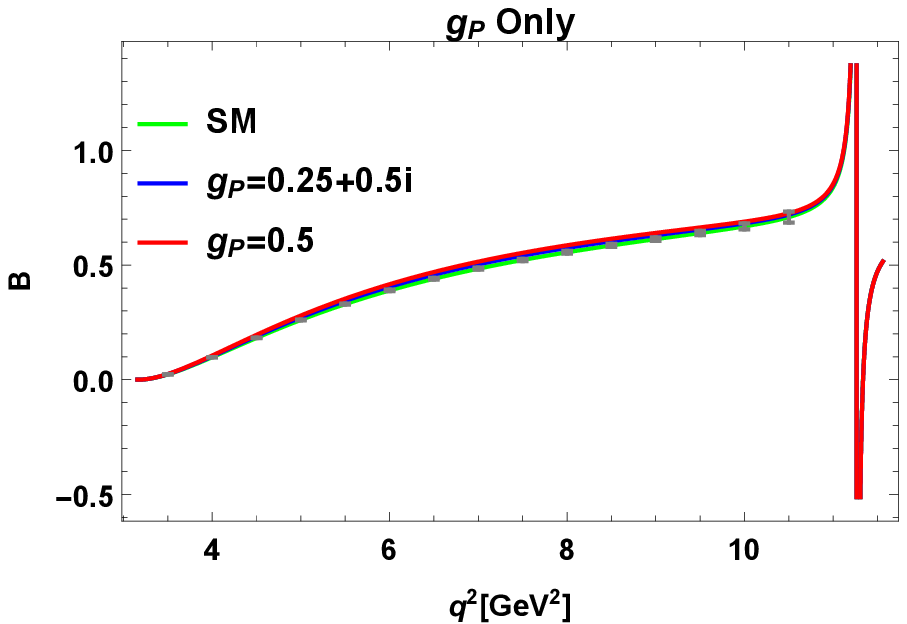}\\[0.3cm]
\end{center}
\begin{adjustwidth}{-0.5cm}{-0.5cm}
\includegraphics[width=5cm]{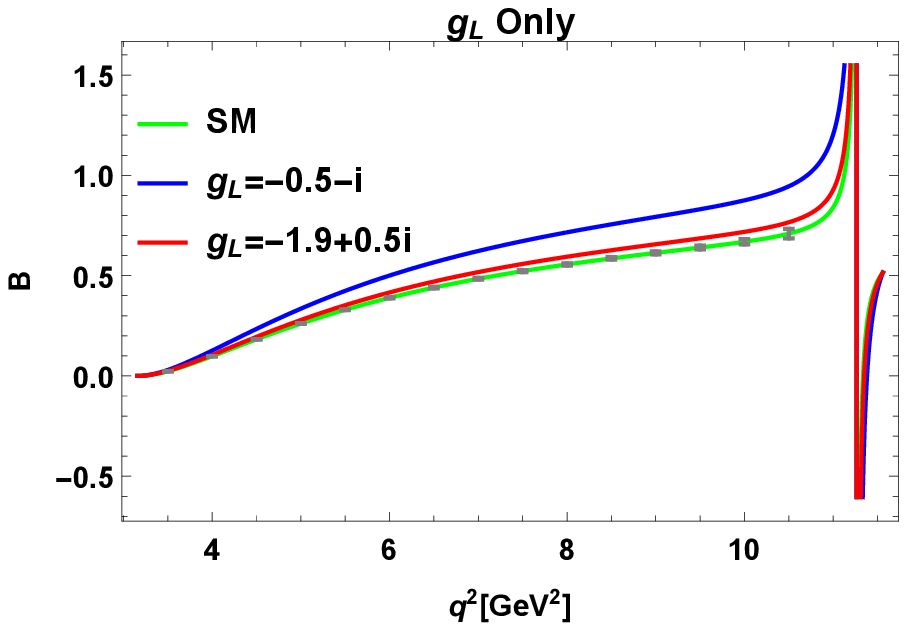}~~~
\includegraphics[width=5cm]{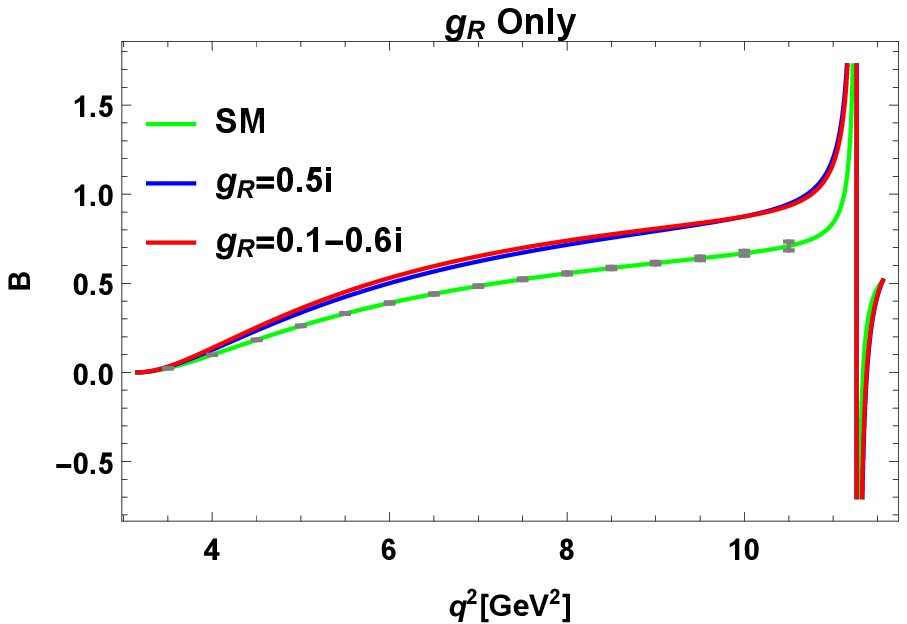}~~~
\includegraphics[width=5cm]{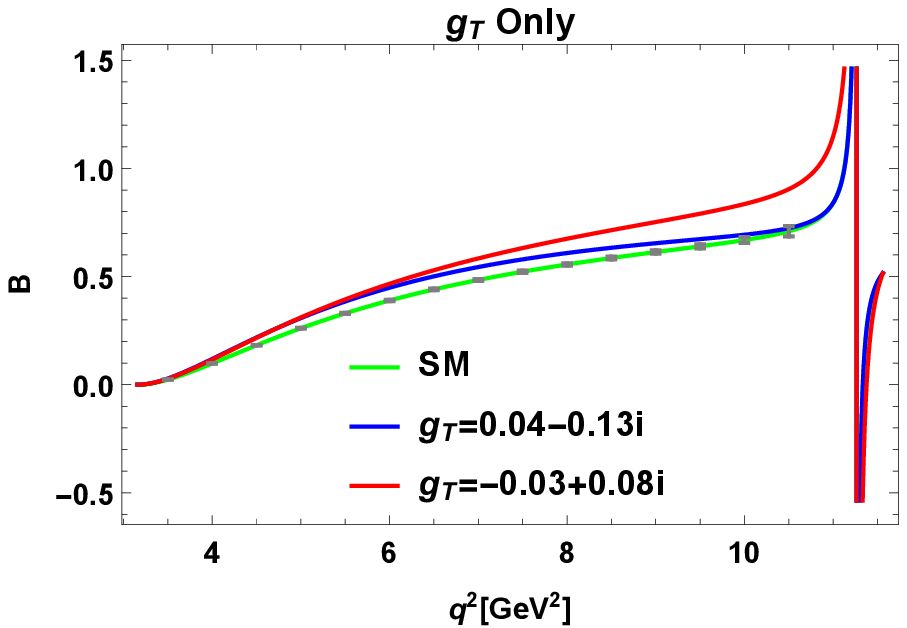}~~~
\end{adjustwidth}
\caption{The effect of individual new-physics couplings on the $B$ ratio, including the QCD $\mathcal{O}(\alpha_s)$ and $1/m_b^2$ correction in the SM contribution only. Each plot shows the observable in the Standard Model and for two allowed values of the new-physics couplings.}
\label{fig:individualcouplingexamples22}
\end{figure}

\begin{figure}
\begin{center}
\includegraphics[width=5cm]{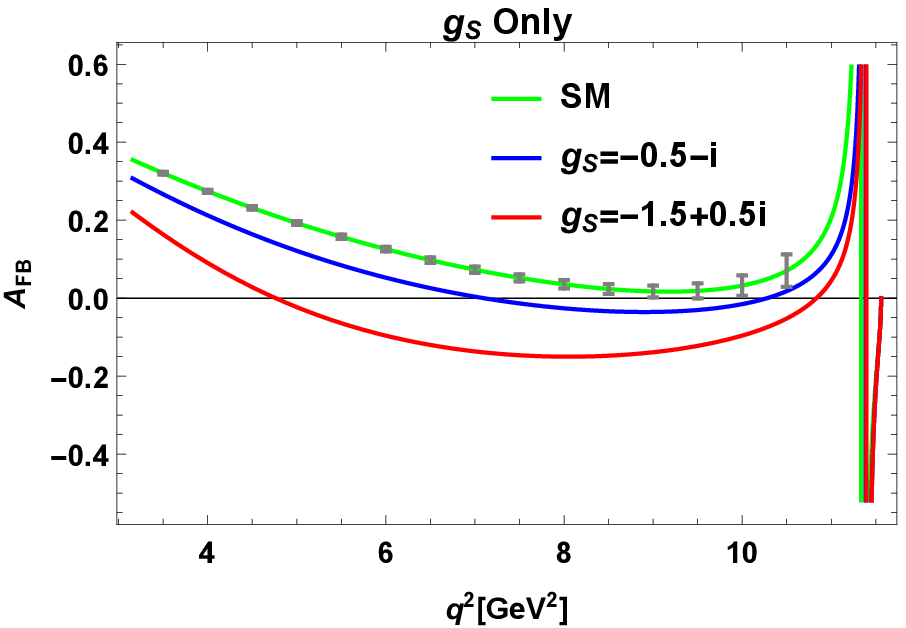}~~~
\includegraphics[width=5cm]{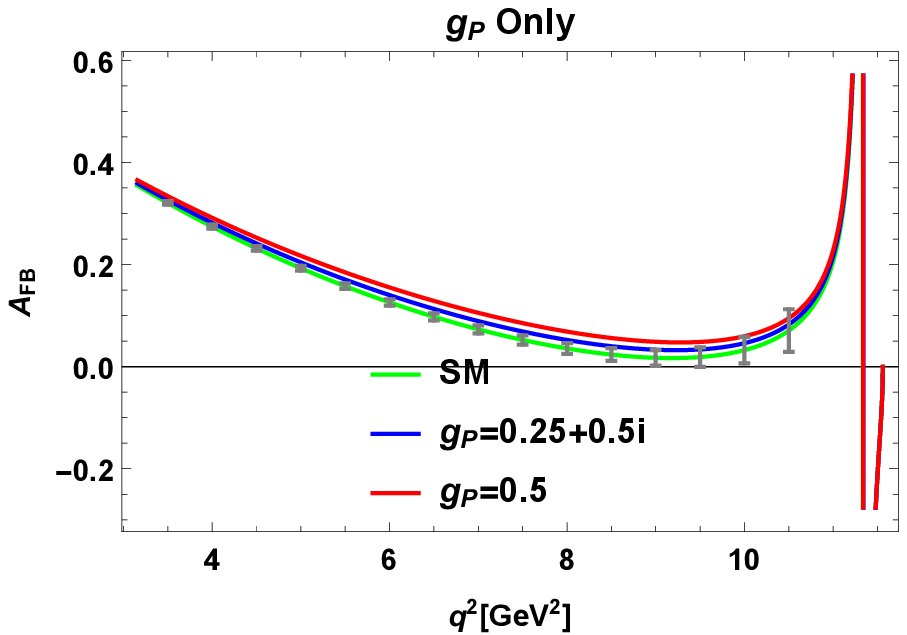}\\[0.3cm]
\end{center}
\begin{adjustwidth}{-0.5cm}{-0.5cm}
\includegraphics[width=5cm]{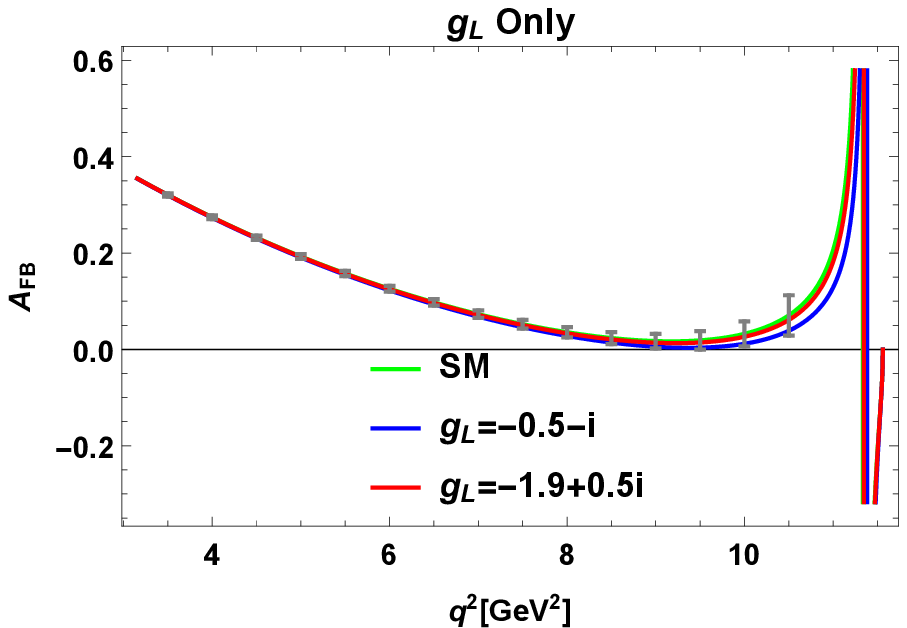}~~~
\includegraphics[width=5cm]{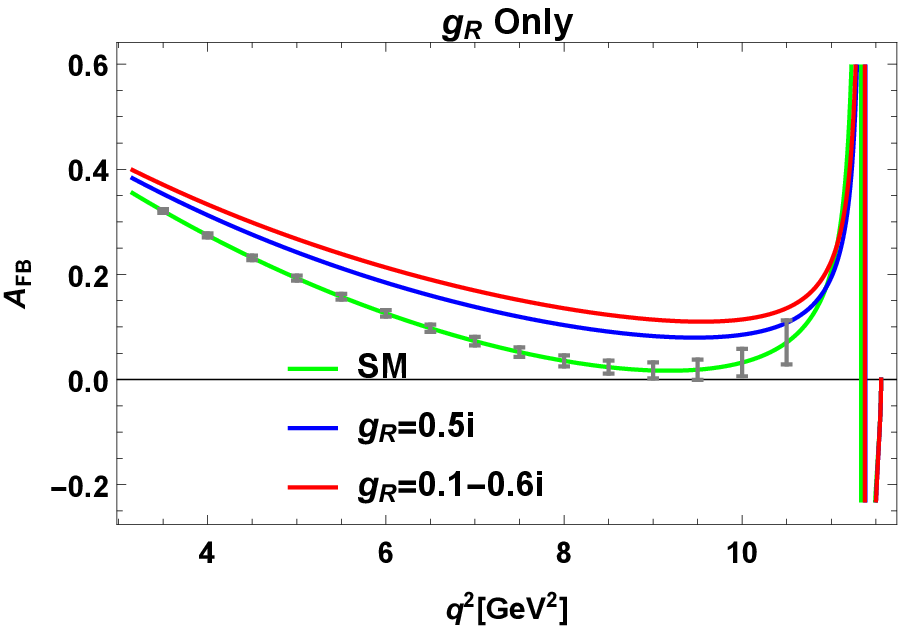}~~~
\includegraphics[width=5cm]{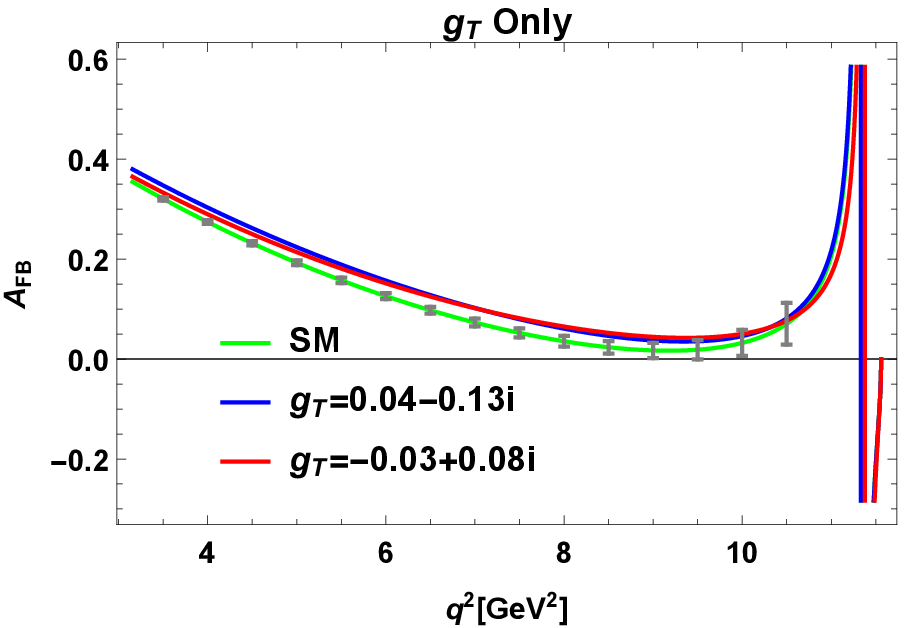}~~~
\end{adjustwidth}
\caption{The effect of individual new-physics couplings on the $\bt$ forward-backward asymmetry $A_{FB}$, including the QCD $\mathcal{O}(\alpha_s)$ and $1/m_b^2$ correction in the SM contribution only. Each plot shows the observable in the Standard Model and for two allowed values of the new-physics couplings.}
\label{fig:individualcouplingexamples2}
\end{figure}

\section{Leptoquark model results}
\label{sec:models}

In this section we introduce leptoquark models that can be the origin of the general couplings in the effective Hamiltonian (\ref{eq1:Lag}). In a recent work \cite{Datta:2017aue} we considered these models in detail. To be self-contained, we briefly describe how these models generate the couplings in the effective Hamiltonian (\ref{eq1:Lag}).\\



In Ref.~\cite{Dumont:2016xpj} several leptoquark models are considered as possible NP scenarios. These models can be grouped as scalar and vector leptoquarks where they can generate scalar $(g_S, g_P)$; vector $(g_L)$; and tensor $(g_T)$ couplings as follows:

\begin{itemize}
\item{ The ${\bm S}_3$ and ${\bm U}_3$ triplet scalar and vector leptoquarks generate the vector coupling $g_L$.}
\item{ The $U_1$ singlet vector leptoquark generates scalar ($g_S, g_P$) and vector ($g_L$) couplings.}
\item{ The $R_2$ doublet scalr leptoquark generates scalar ($g_S, g_P$) and tensor ($g_T$) couplings.}
\item{ The $S_1$ singlet scalar leptoquark generates scalar ($g_S, g_P$), vector ($g_L$) and tensor ($g_T)$ couplings.}
\end{itemize}

The leptoquark Lagrangian generates these couplings in the following way:
\begin{eqnarray}
g_S &=& \frac{\sqrt{2}}{4G_F V_{cb}} \sum_{k=1}^3 V_{k3} 
      \left[ 
      -{2g_{2L}^{kl}g_{2R}^{23*} \over M_{V_2}^2} - {2h_{1L}^{2l}h_{1R}^{k3*} \over M_{U_1}^2} -{g_{1L}^{kl}g_{1R}^{23*} \over 2M_{S_1}^2} - {h_{2L}^{2l}h_{2R}^{k3*} \over 2M_{R_2}^2} \right] ,\\  
g_P &=& \frac{\sqrt{2}}{4G_F V_{cb}}  \sum_{k=1}^3 V_{k3} 
      \left[ -{2g_{2L}^{kl}g_{2R}^{23*} \over M_{V_2}^2} - {2h_{1L}^{2l}h_{1R}^{k3*} \over M_{U_1}^2} + {g_{1L}^{kl}g_{1R}^{23*} \over 2M_{S_1}^2} + {h_{2L}^{2l}h_{2R}^{k3*} \over 2M_{R_2}^2} \right],\\
g_L &=& \frac{\sqrt{2}}{4G_F V_{cb}} \sum_{k=1}^3 V_{k3} 
      \left[ 
      {g_{1L}^{kl}g_{1L}^{23*} \over 2M_{S_1}^2} - {g_{3L}^{kl}g_{3L}^{23*} \over 2M_{{\bm S}_3}^2} + {h_{1L}^{2l}h_{1L}^{k3*} \over M_{U_1}^2} - {h_{3L}^{2l}h_{3L}^{k3*} \over M_{{\bm U}_3}^2}
      \right] \,, \\
g_R &=& 0,\\
g_T &=& \frac{\sqrt{2}}{4G_F V_{cb}} \sum_{k=1}^3 V_{k3} 
      \left[ 
      {g_{1L}^{kl}g_{1R}^{23*} \over 8M_{S_1}^2} - {h_{2L}^{2l}h_{2R}^{k3*} \over 8M_{R_2}^2} 
      \right] \,.
\end{eqnarray}

where $g^{ij}$ and $h^{ij}$ are the leptoquark couplings with $i(j)$ indicating the generation of quarks (leptons) and $M$'s are leptoquark masses with the subscripts corresponding to the leptoquark type. One should run these couplings down to the $b$ quark mass scale as they are defined at the leptoquark mass scale $(\sim 1~TeV)$.  Here $V_{k3}$ corresponds to the CKM matrix element, with $3$ referring to the bottom quark.  We neglect the CKM-suppressed contributions from $k=1$ and $k=2$.

For completeness we just remind the reader that the leptoquark couplings can also be constrained by $b \to s \nu \bar{\nu}$ decays, so we also consider the exclusive $B \to K^{(*)} \nu \bar{\nu}$ decays in our analysis. Following Ref.~\cite{Sakaki:2013bfa}, the $b\to s \nu_j \bar{\nu}_i$ process can be described by the effective Hamiltonian,
\begin{equation}
   H_{eff} = {4G_F \over \sqrt{2}} V_{tb} V_{ts}^* \left[ \left(\delta_{ij}C_L^{(\rm SM)} + C_L^{ij}\right)O_L^{ij} + C_R^{ij}O_R^{ij} \right] \,,
\end{equation}
where the left-handed and right-handed operators are defined as
\begin{equation}
   \begin{split}
      O_L^{ij} =& (\bar{s}_L \gamma^\mu b_L)(\bar{\nu}_{jL} \gamma_\mu \nu_{iL}) \,, \\
      O_R^{ij} =& (\bar{s}_R \gamma^\mu b_R)(\bar{\nu}_{jL} \gamma_\mu \nu_{iL}) \,.
   \end{split}
\end{equation}
The SM Wilson coefficient $C_L^{(\rm SM)}$ receives contributions from the box and the $Z$-penguin diagrams, which yield
\begin{equation}
   C_L^{(\rm SM)} = {\alpha \over 2\pi\sin^2\theta_W}X(m_t^2/M_W^2) \,,
\end{equation}
where the loop function $X(x_t)$ can be found e.g.~in Ref.~\cite{Buras:1998raa}.
Leptoquarks produce contributions to $C_L^{ij}$ which, to leading order, are equal to \cite{Sakaki:2013bfa}
\begin{subequations}
   \label{eq:LQ_coeff_Xsnunu}
   \begin{align}
      C_L^{ij} =& -{1 \over 2\sqrt2 G_F V_{tb} V_{ts}^* } 
      \left[ {g_{1L}^{3i}g_{1L}^{2j*} \over 2M_{S_1}^2} + {g_{3L}^{3i}g_{3L}^{2j*} \over 2M_{S_3}^2} - {2h_{3L}^{2i}h_{3L}^{3j*} \over M_{U_3}^2} \right] \,.
   \end{align}
\end{subequations}
Now we obtain the common coefficients for the $b \to c \tau \bar{\nu}_l$ and $b \to s \nu_\tau \bar{\nu}_l$ processes,

\begin{subequations}
   \begin{align}
      C_L^{l3} =& -{1 \over 2\sqrt2 G_F V_{tb} V_{ts}^* } 
      \left[ {g_{1L}^{3l}g_{1L}^{23*} \over 2M_{S_1}^2} + {g_{3L}^{3l}g_{3L}^{23*} \over 2M_{S_3}^2} - {2h_{3L}^{2l}h_{3L}^{33*} \over M_{U_3}^2} \right] \,.
   \end{align}
\end{subequations}

Hence, for $l=3$ we obtain
\bea
\frac{\mathcal B_K^{{\rm SM} + {\rm NP}}}{\mathcal B_K^{\rm SM}} = \frac{\mathcal B_{K^*}^{{\rm SM} + {\rm NP}}}{\mathcal B_{K^*}^{\rm SM}}  & = & \left | \frac{ 3C_L^{(\rm SM)}  + C_L^{33}}
{3C_L^{(\rm SM)}} \right | ^2 ,\
\eea

while for $l=1,2$ we have

\bea
\frac{\mathcal B_K^{{\rm SM} + {\rm NP}}}{\mathcal B_K^{\rm SM}} = \frac{\mathcal B_{K^*}^{{\rm SM} + {\rm NP}}}{\mathcal B_{K^*}^{\rm SM}}  & = & \left | \frac{  C_L^{l3}}
{3C_L^{(\rm SM)}} \right |^2 .\
\eea 
Now we apply leptoquark models to the inclusive decay $\bt$. 
In leptoquark models in general, we can have all neutrino generations coupled to the $\tau$ lepton as NP effects. We impose the constraints on all the leptoquark couplings simultaneously from the experimental measurements of $R(D)$ and $R(D^*)$ within a 3$\sigma$ confidence level, as well as $\tau_{B_c}$ and $\mathcal{B}(B \to K^{(*)} \nu \bar{\nu})$. Then we substitute the allowed values of the couplings in the calculations of $R_D^{Ratio}$, $R_{D^*}^{Ratio}$, and $R_{X_c}^{Ratio}$ to demonstrate the allowed regions of these observables in the presence of each leptoquark model. The results are presented in Fig. \ref{scatter_plots}.

\begin{figure}
\begin{center}
\includegraphics[width=5cm]{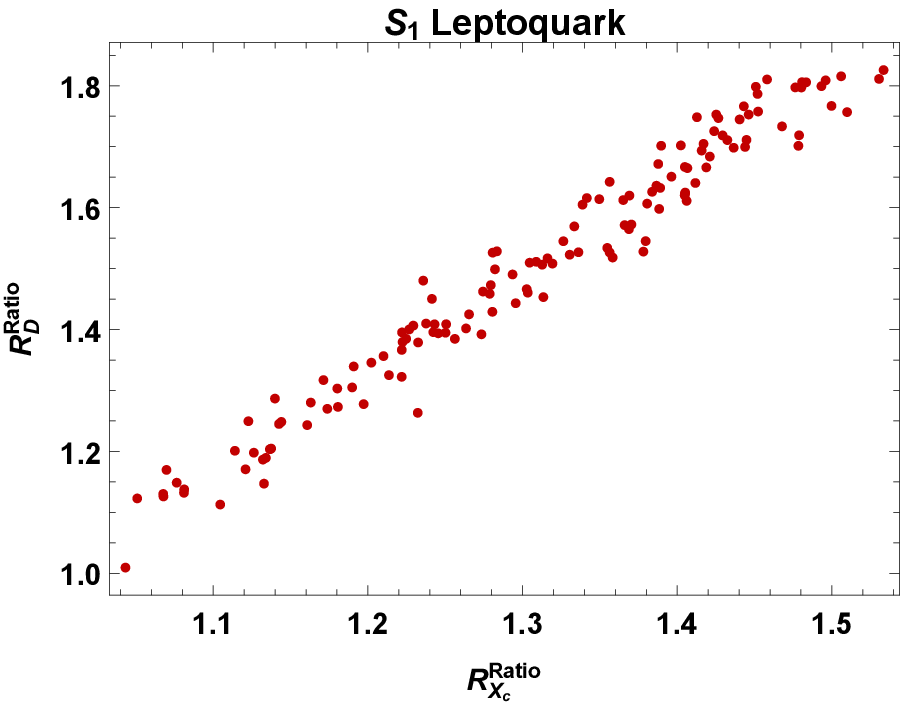}~~~
\includegraphics[width=5cm]{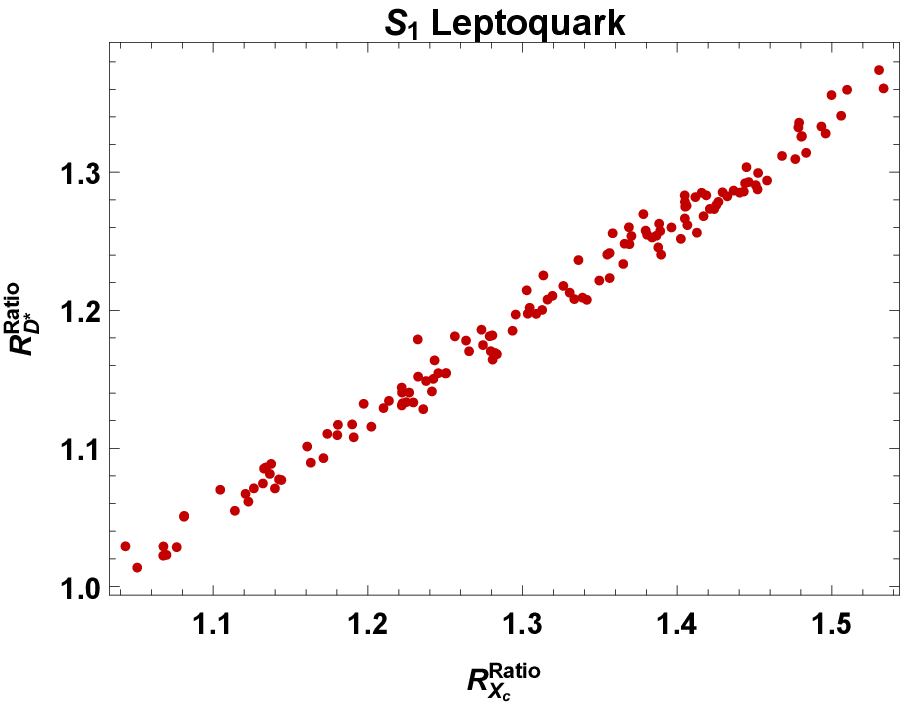}~~~\\
\includegraphics[width=5cm]{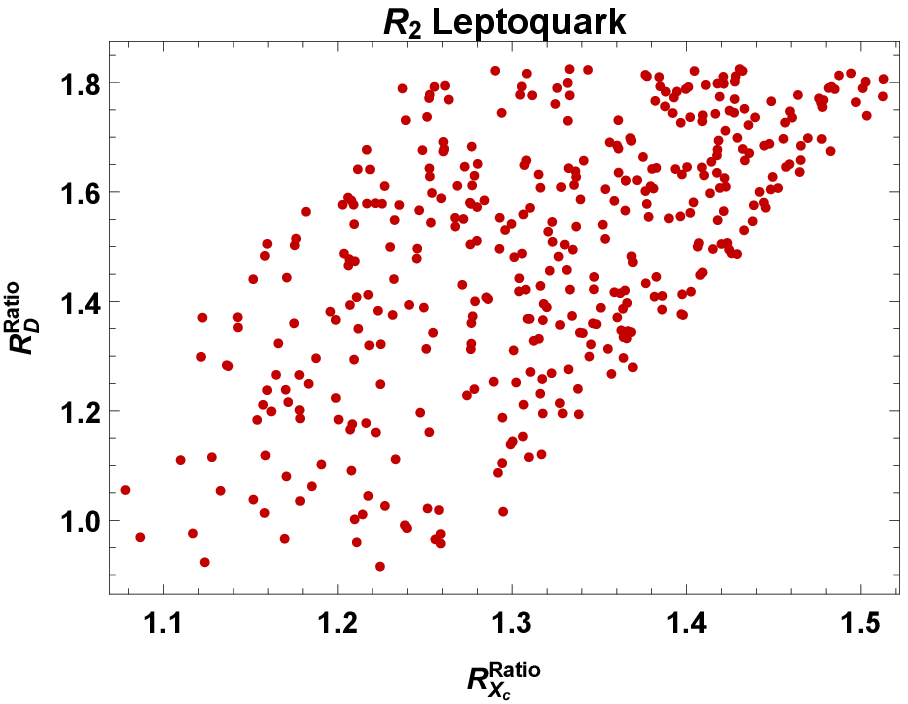}~~~
\includegraphics[width=5cm]{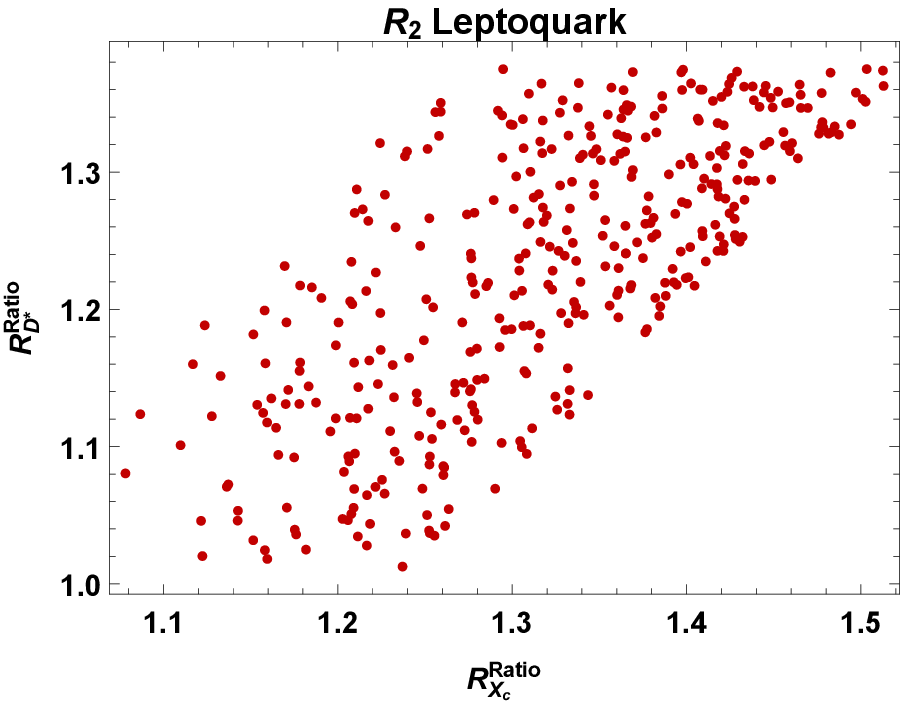}~~~\\
\includegraphics[width=5cm]{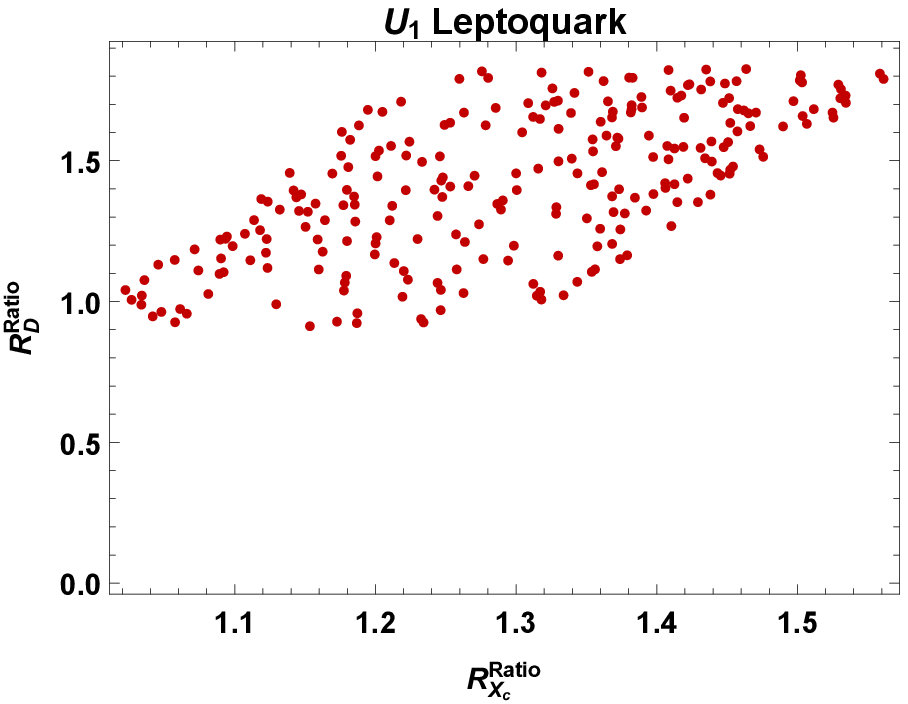}~~~
\includegraphics[width=5cm]{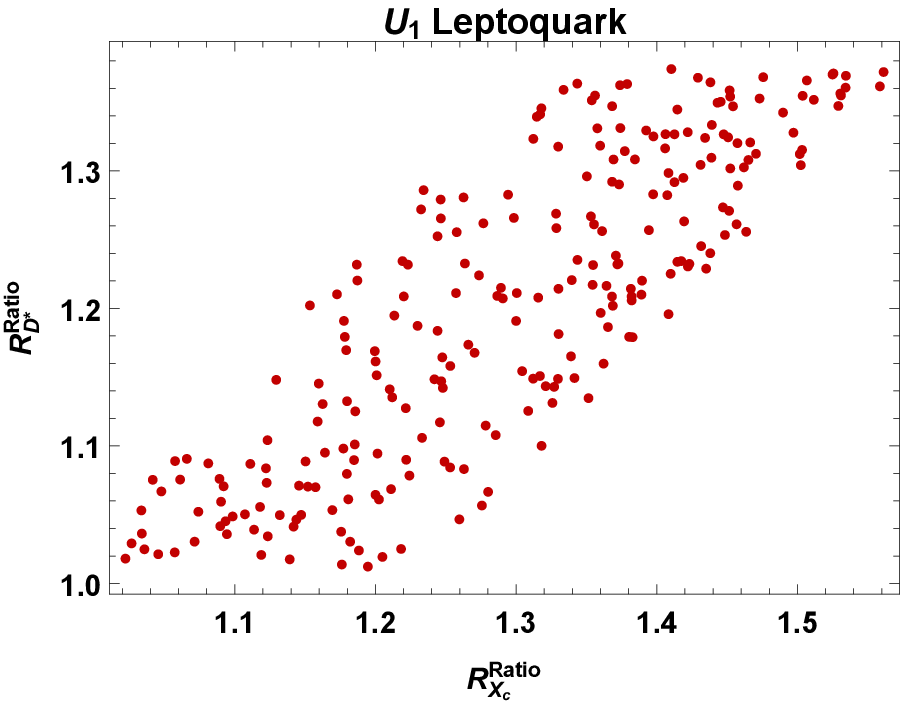}~~~\\
\includegraphics[width=5cm]{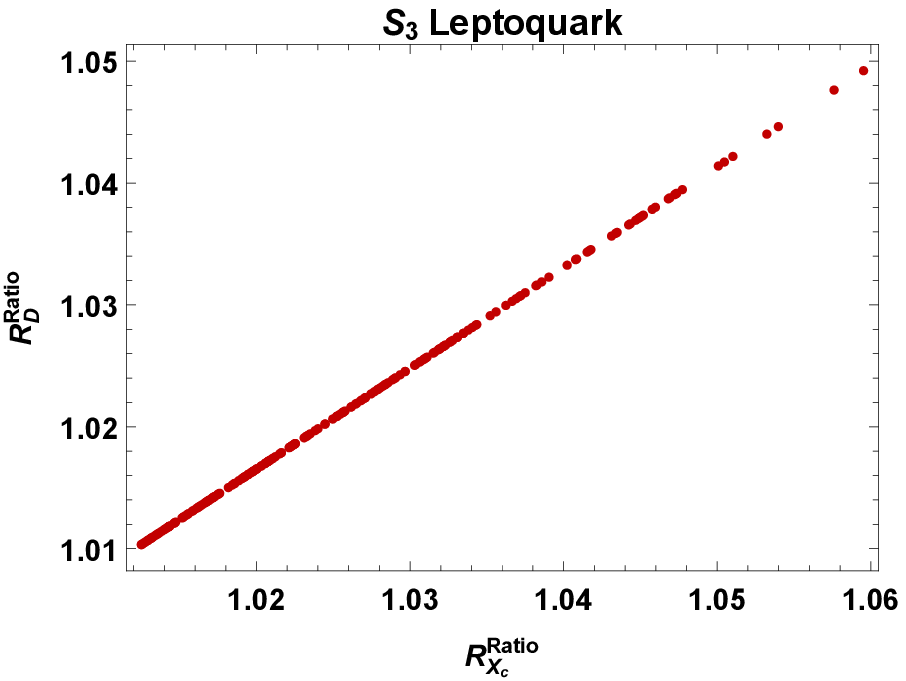}~~~
\includegraphics[width=5cm]{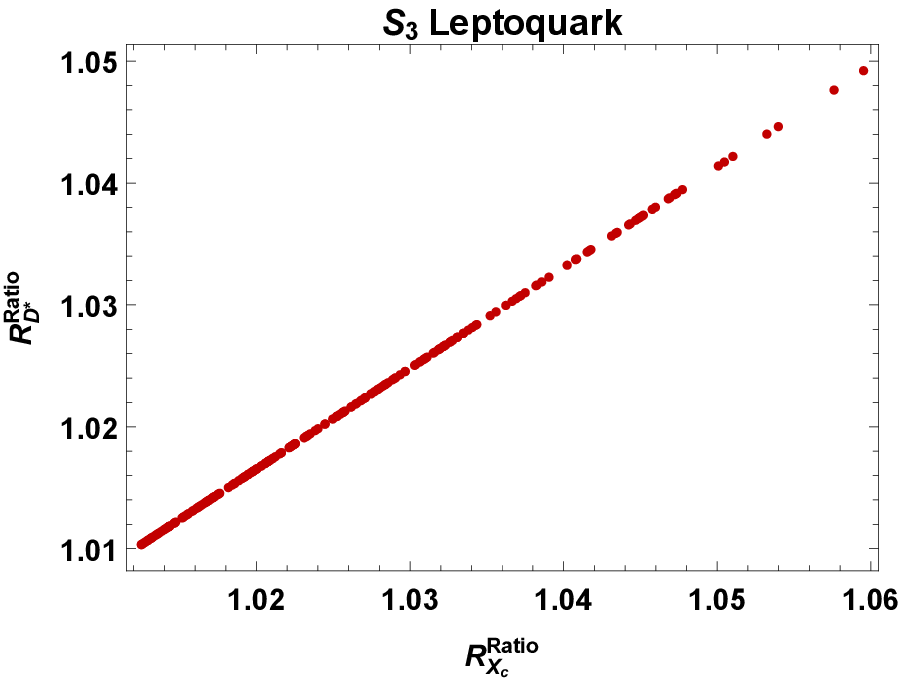}~~~\\
\includegraphics[width=5cm]{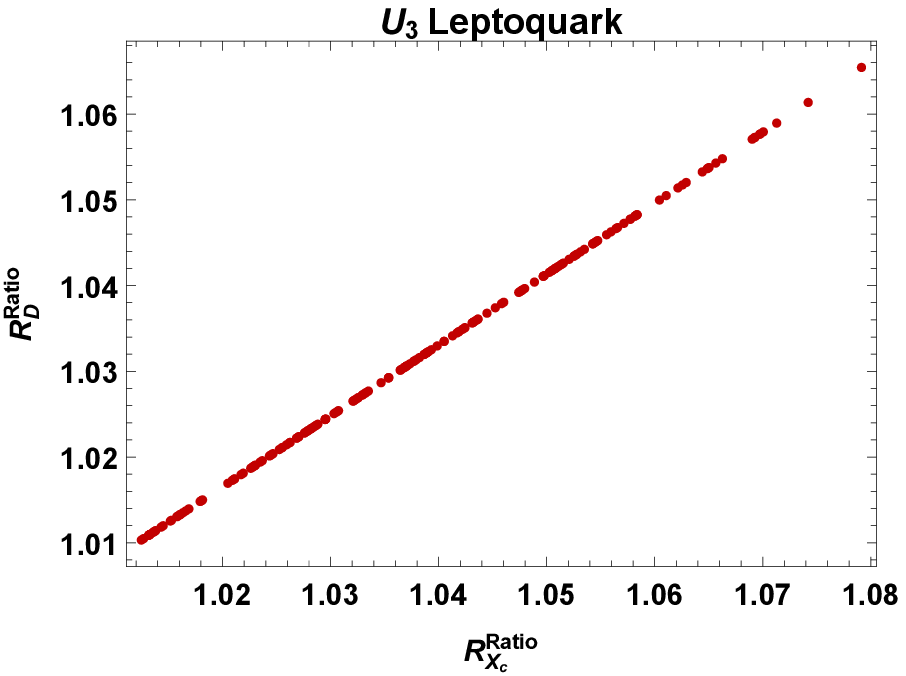}~~~
\includegraphics[width=5cm]{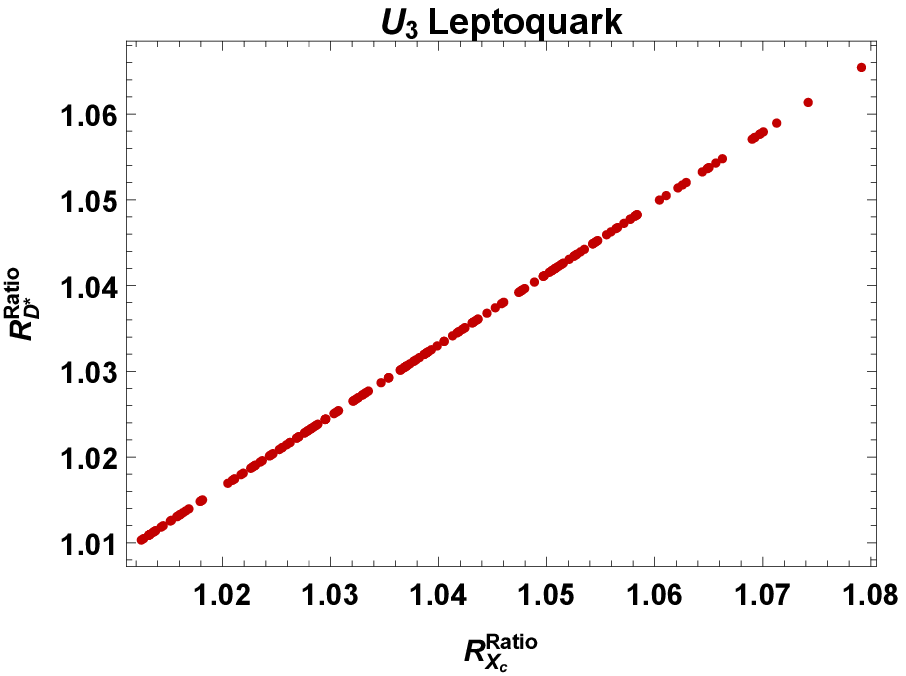}~~~
\end{center}
\caption{The allowed regions in the $R_{X_c}^{Ratio}-R_D^{Ratio}$ and $R_{X_c}^{Ratio}-R_{D^*}^{Ratio}$ planes for each leptoquark model where the couplings are constrained by measurements of $R(D)$ and $R(D^*)$, the branching ratio of $B_c \to \tau^- \bar{\nu}_\tau$, and are consistent with the upper bounds on $\mathcal{B}(B \to K^{(*)}\nu \bar{\nu})$ at $90\%$ C.L. }
\label{scatter_plots}
\end{figure}

Since in leptoquark models in general, there can be multiple NP couplings present (as opposed to model independent scenarios where one coupling at a time is considered), in  Figs. \ref{LQ-shapes1} - \ref{LQ-shapes5} we present the effect of different leptoquark models (${S}_1$, ${R}_2$,  ${U}_1$, ${\bm S}_3$, ${\bm U}_3$) for some allowed values of the model parameters on the inclusive decay $\bt$ observables.
${\bm S}_3$ and ${\bm U}_3$ models are tightly constrained and only  small effects are possible, while other models can have large effects on the considered observables.
This can be seen  in the correlation plots in the $R_{X_c}^{Ratio}-R_{D}^{Ratio}$ and $R_{X_c}^{Ratio}-R_{D^*}^{Ratio}$ planes where in the ${\bm S}_3$ and ${\bm U}_3$ models we see small deviations of the $R$ values from the SM predictions while large deviations are possible with the other leptoquarks.  
 The differential distributions can have different shapes from the SM and $A_{FB}$ can have zero crossings and take negative values for certain leptoquark models. The pattern of deviations from the SM can also be different for the different leptoquark models.
 Hence the careful measurements of these observables can point to the presence of leptoquarks and give clues to their structures. As discussed in the previous section, the $CP$ violation in the inclusive $B$ decay suggested by the imaginary parts of the couplings in the  leptoquark model may be discussed in a separate paper.

\begin{figure}
\begin{center}
\includegraphics[width=5cm]{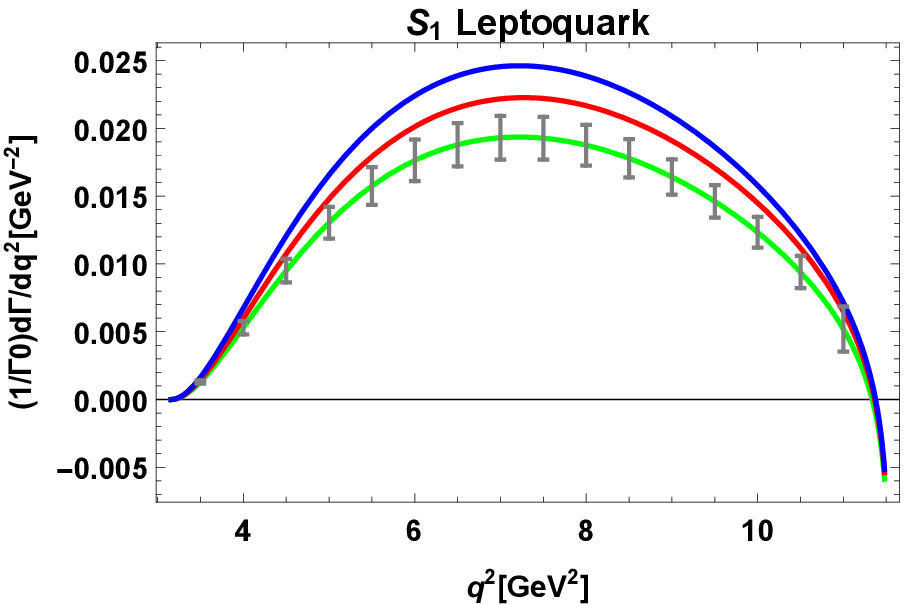}~~~
\includegraphics[width=5cm]{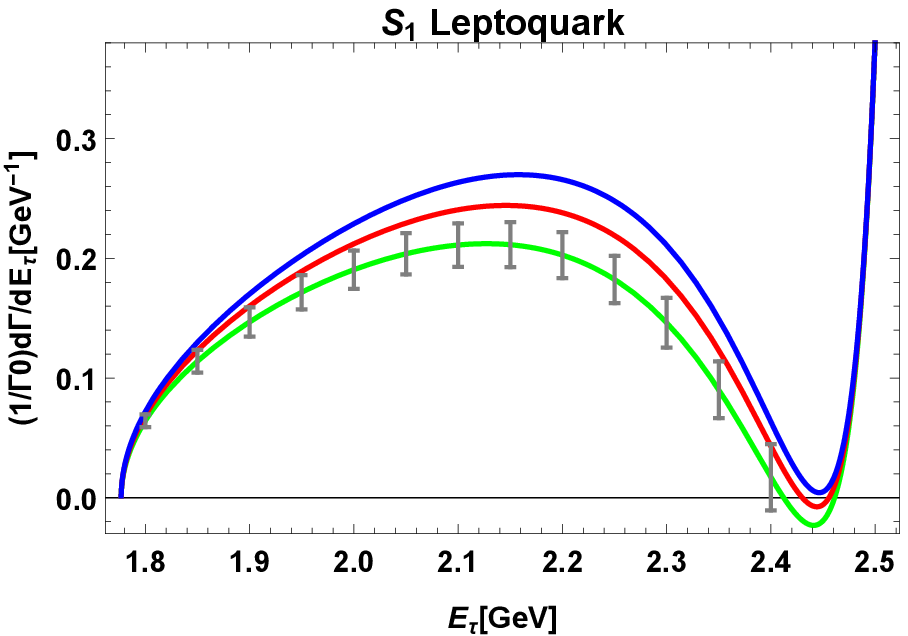}~~~\\
\includegraphics[width=5cm]{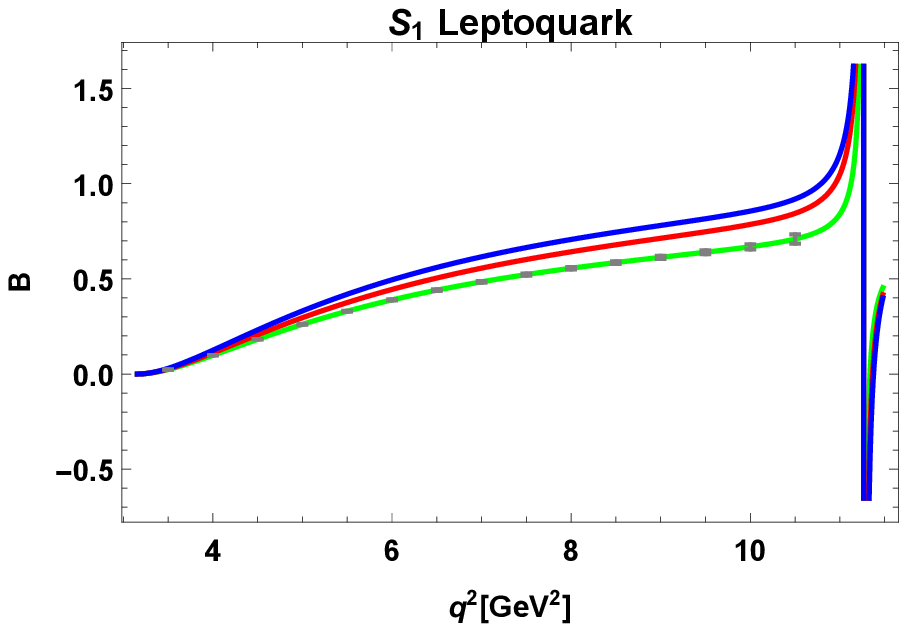}~~~
\includegraphics[width=5cm]{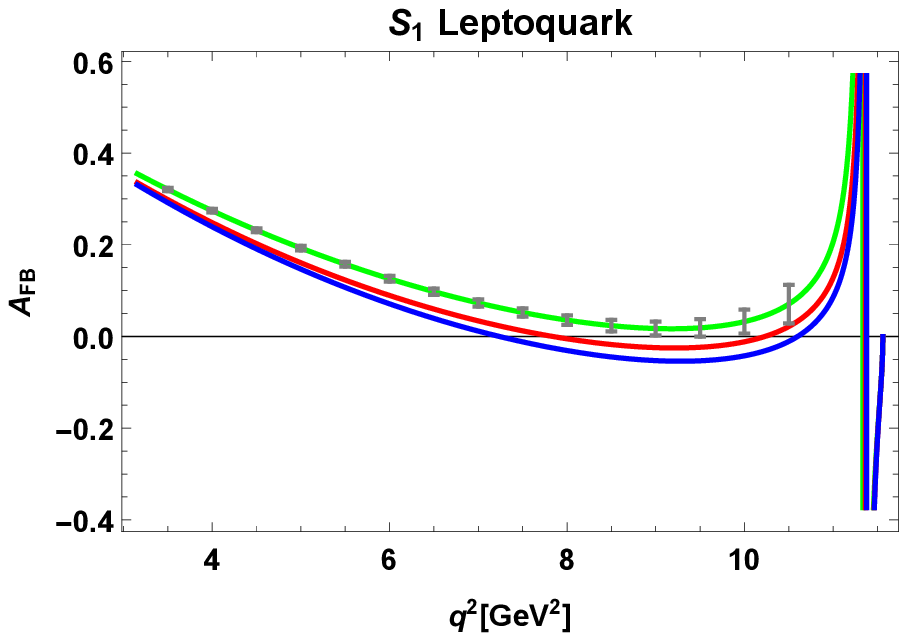}
\end{center}
\caption{The effects of the ${S}_1$ leptoquark model on the differential decay rates $(1/\Gamma_0)d\Gamma/dq^2$, $(1/\Gamma_0)d\Gamma/dE_\tau$; the ratio of differential rates $B$; and the forward-backward asymmetry $(A_{FB})$ of $\bt$. Each plot shows the observable in the Standard Model and for two allowed values of the NP couplings. The red curves correspond to $g_{1L}^{33}g_{1R}^{23*}= 0.203 + 0.121 i$, $g_{1L}^{32}g_{1R}^{23*}= 1.100 - 0.385 i$, $g_{1L}^{31}g_{1R}^{23*}= 0.270 + 0.149 i$, $g_{1L}^{33}g_{1L}^{23*}= -0.015 + 0.014 i$, $g_{1L}^{32}g_{1L}^{23*}= -0.027 - 0.031 i$, $g_{1L}^{31}g_{1L}^{23*}= -0.054 - 0.009 i$, and the blue curves correspond to $g_{1L}^{33}g_{1R}^{23*}= 0.420 - 0.369 i$, $g_{1L}^{32}g_{1R}^{23*}= -0.818 - 0.253 i$, $g_{1L}^{31}g_{1R}^{23*}= 0.711 + 0.761 i$, $g_{1L}^{33}g_{1L}^{23*}= 0.095 + 0.002 i$, $g_{1L}^{32}g_{1L}^{23*}= -0.042 - 0.110 i$, $g_{1L}^{31}g_{1L}^{23*}= -0.003 - 0.022 i$, while the green curves correspond to the Standard Model.}
\label{LQ-shapes1}
\end{figure}

\begin{figure}
\begin{center}
\includegraphics[width=5cm]{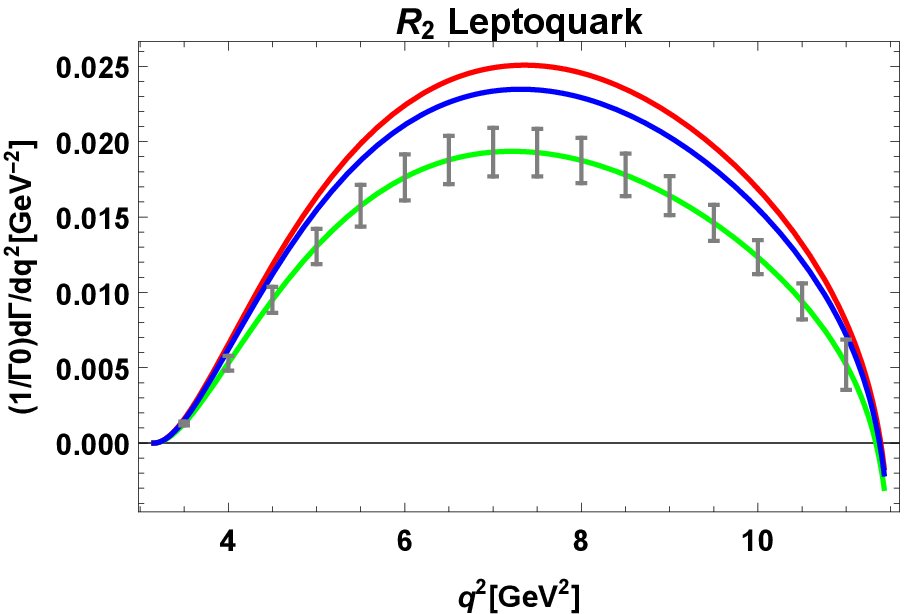}~~~
\includegraphics[width=5cm]{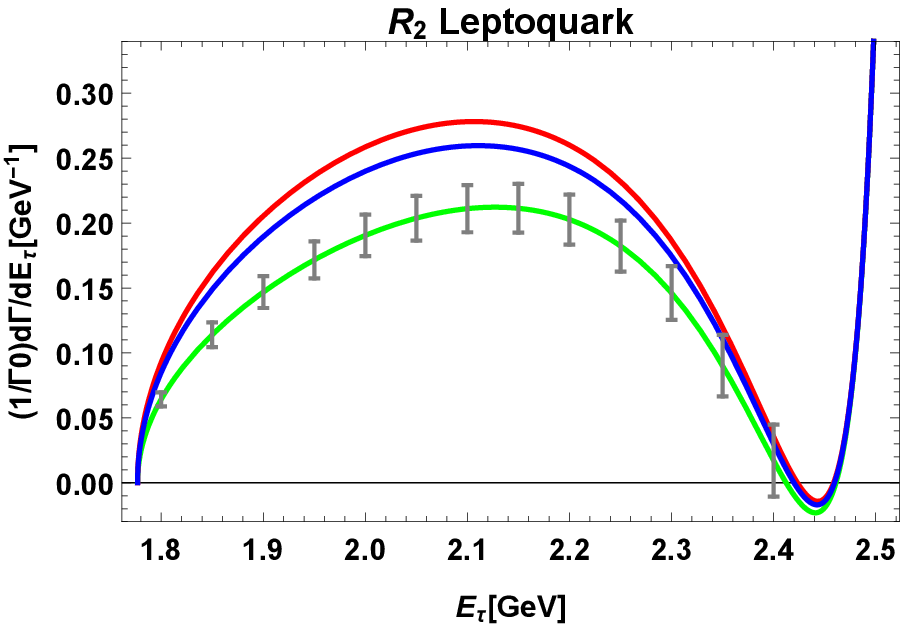}~~~\\
\includegraphics[width=5cm]{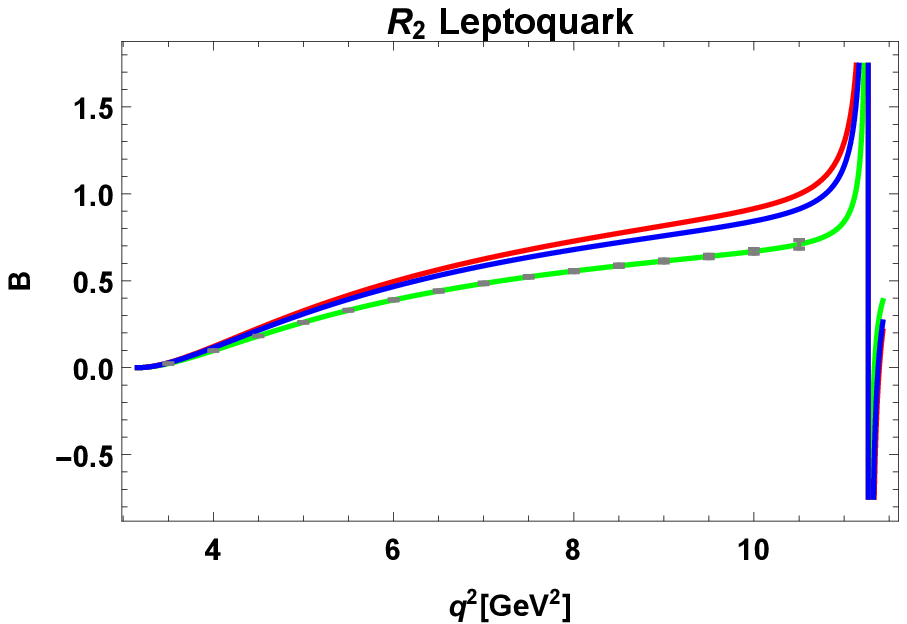}~~~
\includegraphics[width=5cm]{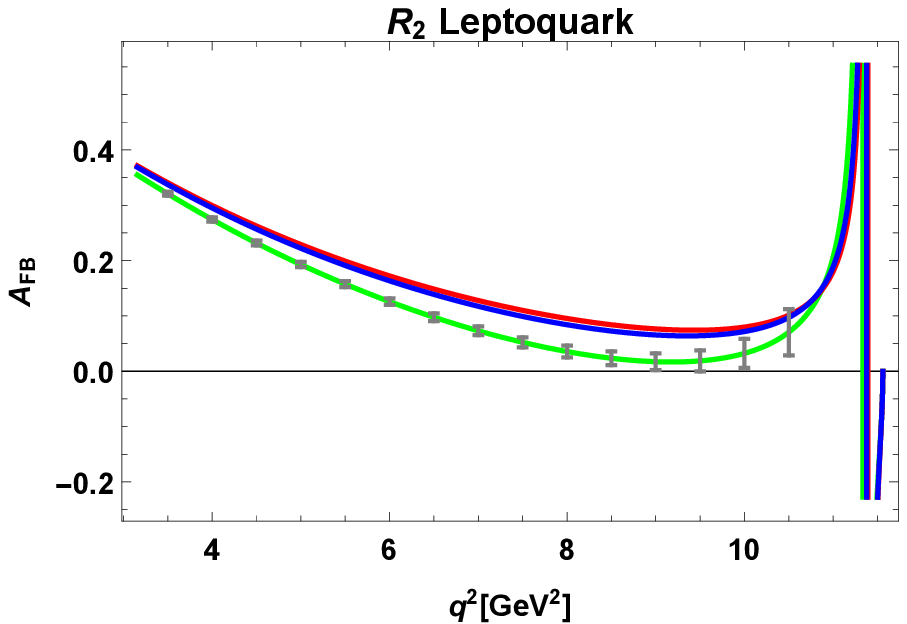}
\end{center}
\caption{The effects of the ${R}_2$ leptoquark model on the differential decay rates $(1/\Gamma_0)d\Gamma/dq^2$, $(1/\Gamma_0)d\Gamma/dE_\tau$; the ratio of differential rates $B$; and the forward-backward asymmetry $(A_{FB})$ of $\bt$. Each plot shows the observable in the Standard Model and for two allowed values of the NP couplings. The red curves correspond to $h_{2L}^{23}h_{2R}^{33*}=0.106 - 0.958 i$, $h_{2L}^{22}h_{2R}^{33*}=-0.218 - 0.546 i$, $h_{2L}^{21}h_{2R}^{33*}=0.493 - 0.134 i$, and the blue curves correspond to $h_{2L}^{23}h_{2R}^{33*}=-0.141 + 0.104 i$, $h_{2L}^{22}h_{2R}^{33*}=-0.814 - 0.647 i$, $h_{2L}^{21}h_{2R}^{33*}=-0.324 - 0.140 i$, respectively, while the green curves correspond to the Standard Model.}
\label{LQ-shapes2}
\end{figure}

\begin{figure}
\begin{center}
\includegraphics[width=5cm]{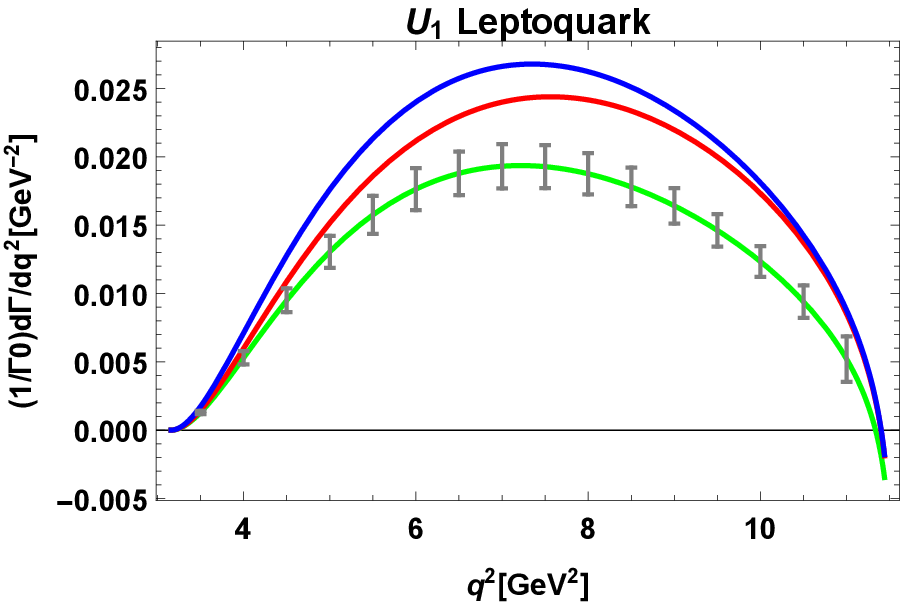}~~~
\includegraphics[width=5cm]{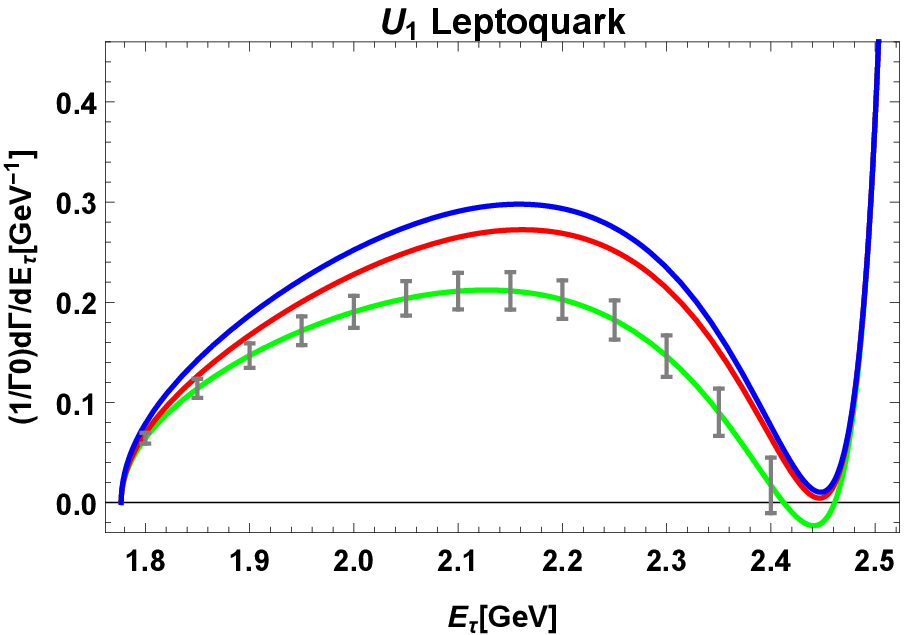}~~~\\
\includegraphics[width=5cm]{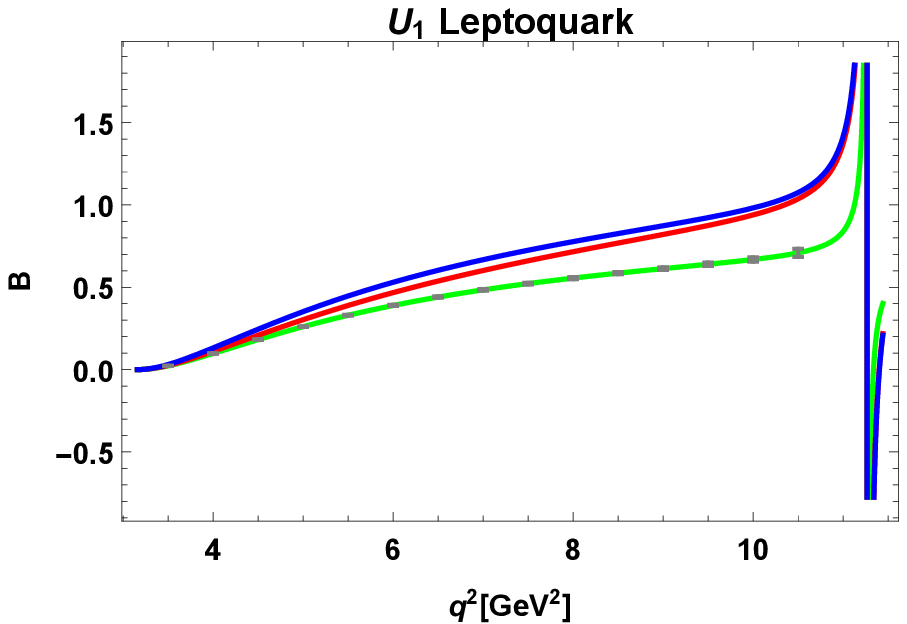}~~~
\includegraphics[width=5cm]{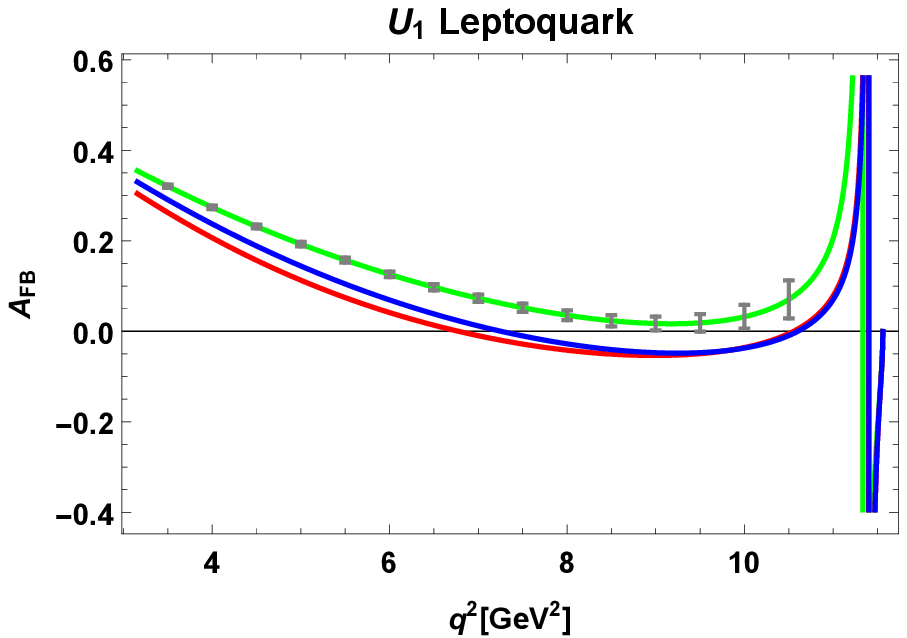}
\end{center}
\caption{The effects of the ${U}_1$ leptoquark model on the differential decay rates $(1/\Gamma_0)d\Gamma/dq^2$, $(1/\Gamma_0)d\Gamma/dE_\tau$; the ratio of differential rates $B$; and the forward-backward asymmetry $(A_{FB})$ of $\bt$. Each plot shows the observable in the Standard Model and for two allowed values of the NP couplings. The red curves correspond to $h_{1L}^{23}h_{1R}^{33*}=-0.127 - 0.395 i$, $h_{1L}^{22}h_{1R}^{33*}=0.077 + 0.043 i$, $h_{1L}^{21}h_{1R}^{33*}=-0.040 + 0.034 i$, $h_{1L}^{23}h_{1L}^{33*}=-1.523 - 0.394 i$, $h_{1L}^{22}h_{1L}^{33*}=0.247 + 0.473 i$, $h_{1L}^{21}h_{1L}^{33*}=0.226 + 1.261 i$, and the blue curves correspond to $h_{1L}^{23}h_{1R}^{33*}=0.017 - 0.028 i$, $h_{1L}^{22}h_{1R}^{33*}=-0.115 + 0.017 i$, $h_{1L}^{21}h_{1R}^{33*}=-0.238 - 0.041 i$, $h_{1L}^{23}h_{1L}^{33*}=-1.22 + 0.301 i$, $h_{1L}^{22}h_{1L}^{33*}=0.730 - 0.039 i$, $h_{1L}^{21}h_{1L}^{33*}=-1.327 + 0.357 i$, respectively, while the green curves correspond to the Standard Model.}
\label{LQ-shapes3}
\end{figure}

\begin{figure}
\begin{center}
\includegraphics[width=5cm]{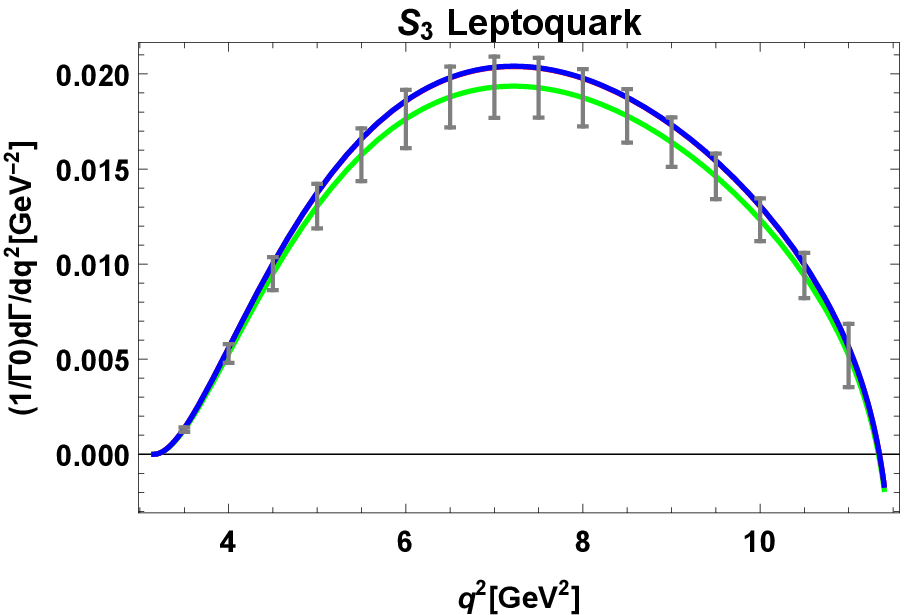}~~~
\includegraphics[width=5cm]{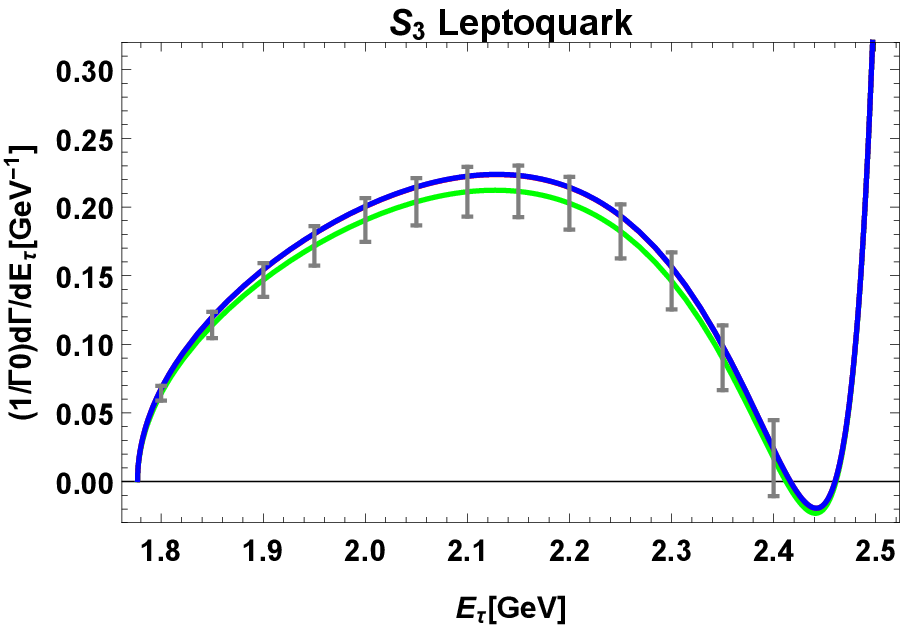}~~~\\
\includegraphics[width=5cm]{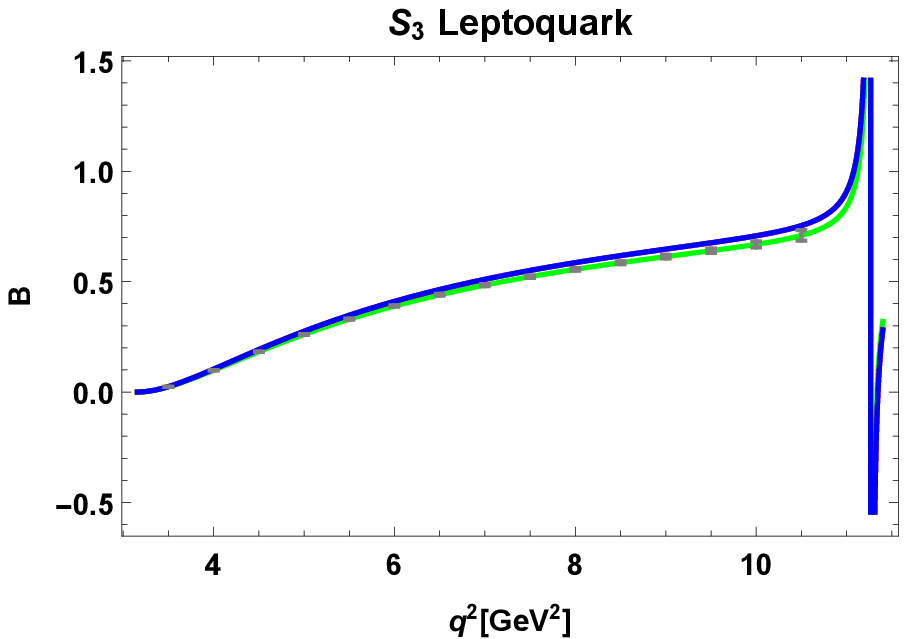}~~~
\includegraphics[width=5cm]{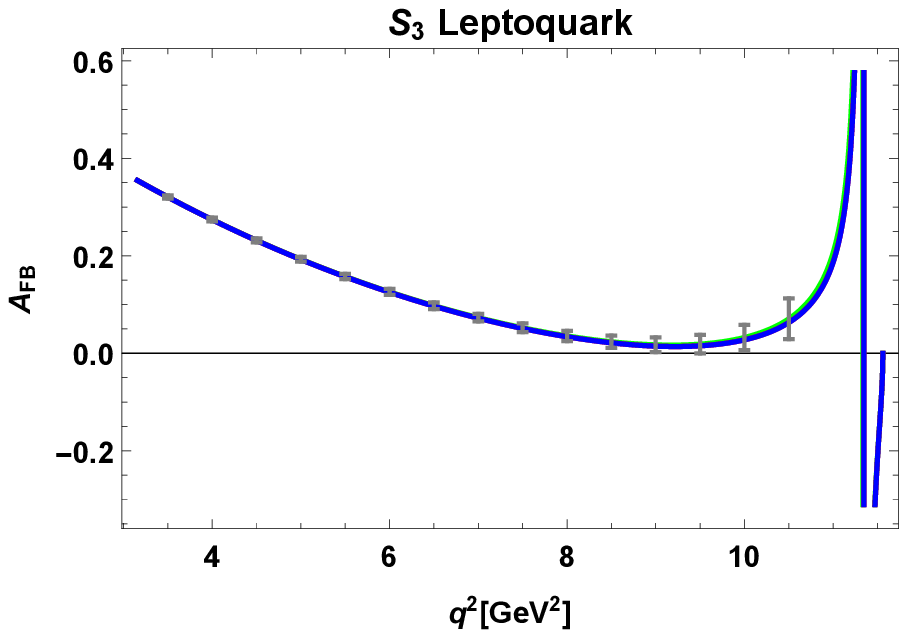}
\end{center}
\caption{The effects of the ${\bm S}_3$ leptoquark model on the differential decay rates $(1/\Gamma_0)d\Gamma/dq^2$, $(1/\Gamma_0)d\Gamma/dE_\tau$; the ratio of differential rates $B$; and the forward-backward asymmetry $(A_{FB})$ of $\bt$. Each plot shows the observable in the Standard Model and for two allowed values of the NP couplings. The red curves correspond to $g_{3L}^{33}g_{3L}^{23*}=-0.062 - 0.028 i$, $g_{3L}^{32}g_{3L}^{23*}=0.031 - 0.005 i$, $g_{3L}^{31}g_{3L}^{23*}=0.013 - 0.003 i$, and the blue curves correspond to $g_{3L}^{33}g_{3L}^{23*}=-0.062 - 0.028 i$, $g_{3L}^{32}g_{3L}^{23*}=0.003 - 0.031 i$, $g_{3L}^{31}g_{3L}^{23*}=0.052 - 0.054 i$, respectively, while the green curves correspond to the Standard Model.}
\label{LQ-shapes4}
\end{figure}

\begin{figure}
\begin{adjustwidth}{-0.5cm}{-0.5cm}
\begin{center}
\includegraphics[width=5cm]{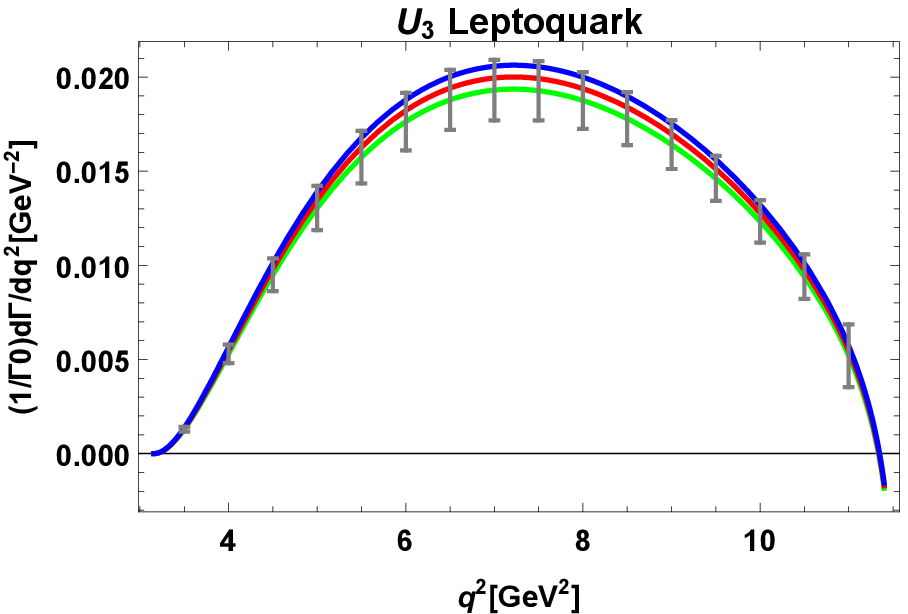}~~~
\includegraphics[width=5cm]{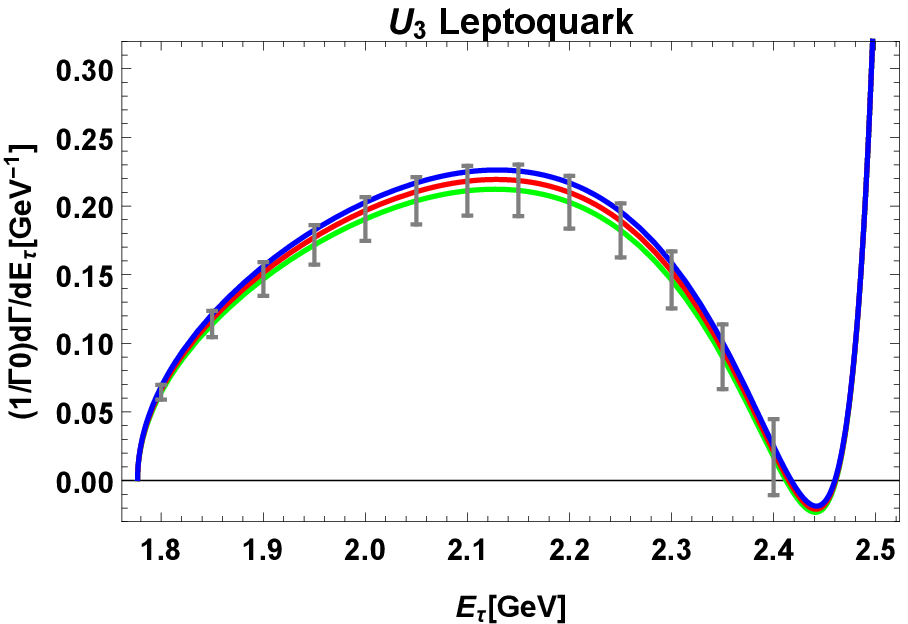}~~~\\
\includegraphics[width=5cm]{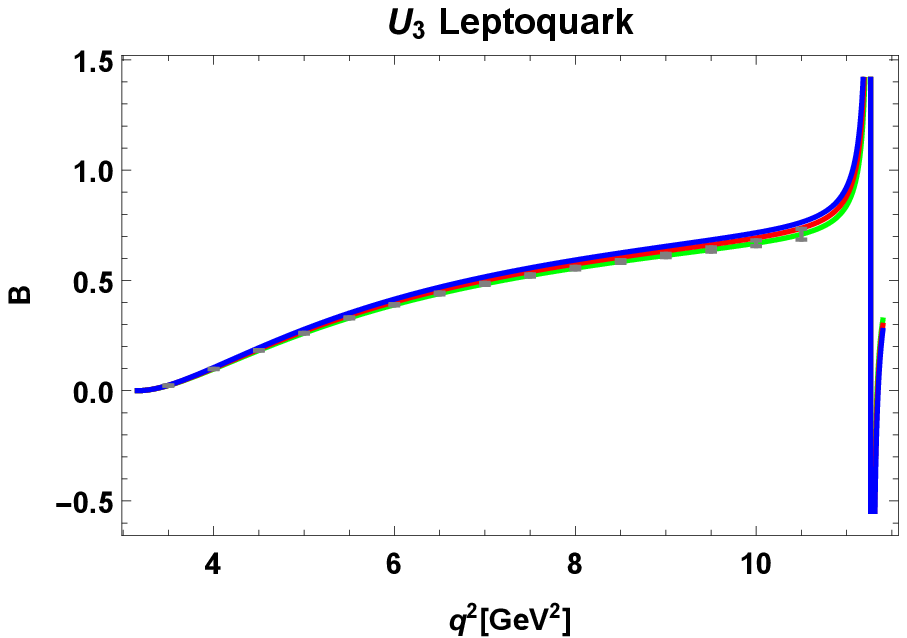}~~~
\includegraphics[width=5cm]{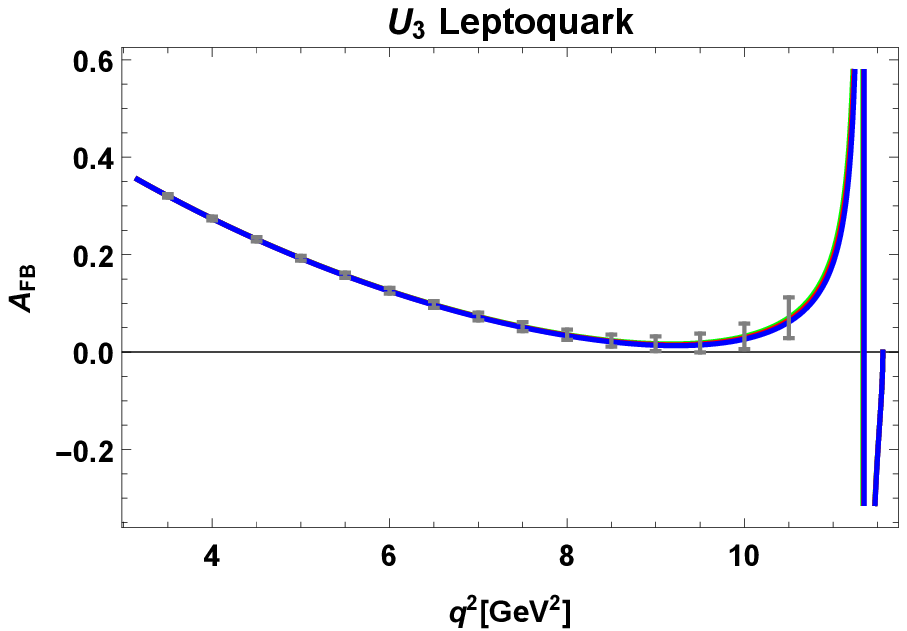}
\end{center}
\end{adjustwidth}
\caption{The effects of the ${\bm U}_3$ leptoquark model on the differential decay rates $(1/\Gamma_0)d\Gamma/dq^2$, $(1/\Gamma_0)d\Gamma/dE_\tau$; the ratio of differential rates $B$; and the forward-backward asymmetry $(A_{FB})$ of $\bt$. Each plot shows the observable in the Standard Model and for two allowed values of the NP couplings. The red curves correspond to $h_{3L}^{23}h_{3L}^{33*}=-0.019 + 0.002 i$, $h_{3L}^{22}h_{3L}^{33*}=0.011 - 0.007 i$, $h_{3L}^{21}h_{3L}^{33*}=0.026 - 0.012 i$, and the blue curves correspond to $h_{3L}^{23}h_{3L}^{33*}=-0.037 + 0.005 i$, $h_{3L}^{22}h_{3L}^{33*}=0.015 + 0.002 i$, $h_{3L}^{21}h_{3L}^{33*}=-0.003 - 0.019 i$, respectively, while the green curves correspond to the Standard Model.}
\label{LQ-shapes5}
\end{figure}

\section{Conclusions}
Recent measurements of $\RD$ that show large deviation from the SM might be providing hints of lepton nonuniversal NP. The underlying transition in these decays $\bctaunutau$ can also be probed in other decays and in this paper
we consider one such process which is the inclusive decay $\bt$. Inclusive decays  suffer from less hadronic uncertainties than exclusive decays and so these decays offer good tests of the SM. In this work we considered NP effects in the 
inclusive decay $\bt$ with the NP parameters constrained by the $\RD$ measurements. We first adopted a model independent approach where the NP is expressed in terms of higher dimensional operators with various Lorentz structures. Considering one
NP operator at a time we considered the effect of NP on the inclusive decay. In the SM the inclusive decays were calculated to 
 perturbative $\mathcal{O}(\alpha_s)$ and $1/m_b^2$ power corrections. Several observables including rates as well as differential distributions were discussed with a particular focus on the ratio of rates
 $R(X_c)= {\mathcal{B}[\bt] \over \mathcal{B}[\bl]}$.  ALEPH has a measurement of $\btq$ which we converted into a measurement of $\bt$ under certain assumptions. Using this as an input we showed that this measurement further constrained the NP couplings
 introduced to address the $\RD$ anomalies. Not including the ALEPH measurement we found that large deviations from the SM in   
 $R(X_c)$ are possible with the present $\RD$ measurements. This highlights the importance of a precise measurement of the inclusive rate as a sensitive probe of NP.
 We then considered  explicit models of NP with leptoquarks and for various models of leptoquarks studied their effects in the inclusive decay. We found that large deviations are possible in certain models of leptoquarks and the patterns of these deviations are different for the different models. Therefore, careful measurements in the inclusive decay can not only point to the presence of leptoquarks but can give clues about their  structure.

\label{sec:conclusion}

\bigskip
\noindent
{\bf Acknowledgments} :
This work was financially supported by the National Science Foundation under Grant No.\ 
PHY-1414345. A. D. acknowledges the hospitality of the Department of Physics and
Astronomy, University of Hawaii and the University of California, Irvine where the work was done.
S. K.  acknowledges the hospitality of the Department of Physics and Astronomy
at the University of California, Irvine.

\clearpage

\appendix

\section{Helicity Amplitudes}
\label{appendix:Helicity}

In general for the process $\bt$, the scalar-type, vector/axial-vector-type, and tensor-type hadronic helicity amplitudes are defined as
\bea
H^{SP}_{\lambda_{c},\lambda=0}&=&H^S_{\lambda_{c},\lambda=0}+H^P_{\lambda_{c},\lambda=0}, \nn\\
H^S_{\lambda_{c},\lambda=0}&=&g_S\bra{X_c}\bar{c} b\ket{B},\nn\\
H^P_{\lambda_{c},\lambda=0}&=&g_P \bra{X_c} \bar{c}\gamma_5 b\ket{B},
\eea
\bea
H^{VA}_{\lambda_{c},\lambda}&=&H^V_{\lambda_{c},\lambda}-H^A_{\lambda_{c},\lambda}, \nn\\
H^V_{\lambda_{c},\lambda}&=&(1+g_L+g_R)\,\epsilon^{*\mu}(\lambda)\bra{X_c}\bar{c}\gamma_{\mu} b\ket{B}, \nn\\
H^A_{\lambda_{c},\lambda}&=&(1+g_L-g_R)\,\epsilon^{*\mu}(\lambda)\bra{X_c}\bar{c}\gamma_{\mu}\gamma_5 b\ket{B},
\eea
and
\bea
H^{(T){\lambda_{b}}}_{\lambda_{c},\lambda ,\lambda^{\prime}}&=&H^{(T1){\lambda_{b}}}_{\lambda_{c},\lambda ,\lambda^{\prime}}-H^{(T2){\lambda_{b}}}_{\lambda_{c},\lambda ,\lambda^{\prime}}, \nn\\
H^{(T1){\lambda_{b}}}_{\lambda_{c},\lambda ,\lambda^{\prime}}&=&g_T\: \epsilon^{*\mu}(\lambda)\epsilon^{*\nu}(\lambda^{\prime})\bra{X_c}\bar{c}i\sigma_{\mu \nu} b\ket{B},\nn\\
H^{(T2){\lambda_{b}}}_{\lambda_{c},\lambda ,\lambda^{\prime}}&=&g_T\:\epsilon^{*\mu}(\lambda)\epsilon^{*\nu}(\lambda^{\prime})\bra{X_c}\bar{c}i\sigma_{\mu \nu}\gamma_5 b\ket{B},
\eea
where $\epsilon^{\mu}$ is the polarization vector of the virtual vector boson.
The leptonic amplitudes are defined as
\bea
L^{\lambda_\tau}&=&\bra{\tau\bar{\nu}_\tau}\bar{\tau} (1-\gamma_5)\nu_\tau\ket{0}, \nn\\
L^{\lambda_\tau}_{\lambda}&=&\epsilon^\mu (\lambda)\bra{\tau\bar{\nu}_\tau}\bar{\tau}\gamma_\mu (1-\gamma_5)\nu_\tau\ket{0}, \nn\\
L^{\lambda_\tau}_{\lambda ,\lambda^{\prime}}&=&-i\epsilon^\mu (\lambda)\epsilon^\nu (\lambda^\prime)\bra{\tau\bar{\nu}_\tau}\bar{\tau}\sigma_{\mu \nu} (1-\gamma_5)\nu_\tau\ket{0}.
\eea
When we consider the process as a free quark decay, we simply use the quark spinors without hadronic expectation values. So the matrix elements for the hadronic vector and axial vector currents will become  
\bea
\bra{X_c}\bar{c}\gamma^\mu  b\ket{B}&\to &\bar{u}_{c}\gamma^\mu u_{b} \label{eq:VFF}, \\
\bra{X_c}\bar{c}\gamma^\mu \gamma_5 b\ket{B}&\to &\bar{u}_{c}\gamma^\mu \gamma_5 u_{b}, \label{eq:AFF}
\eea
for the scalar and pseudoscalar currents 
\bea
\bra{X_c}\bar{c} b\ket{B} & \to & \bar{u}_c u_b,\nonumber\\
\bra{X_c}\bar{c}\gamma_5 b\ket{B} & \to & \bar{u}_c \gamma_5 u_b,
\eea
and for the tensor currents 
\bea
\bra{X_c}\bar{c}i\sigma^{\mu\nu} b\ket{B}& \to & \bar{u}_{c}i \sigma^{\mu \nu} u_{b}, \nonumber \\ 
\bra{X_c}\bar{c}i\sigma^{\mu\nu}\gamma_5 b\ket{B}& \to & \bar{u}_{c}i \sigma^{\mu \nu} \gamma_5 u_{b}. \nonumber \\\label{eq:TFF}
\eea

The hadronic and leptonic helicity amplitudes of the process $b\to c\tau^-\bar{\nu}_\tau$ in the presence of scalar and pseudoscalar, vector and axial-vector, and tensor NP operators are below.

\subsection{Hadronic helicity amplitudes}

Below, we present only the nonvanishing hadronic helicity amplitudes. The scalar and pseudoscalar helicity amplitudes associated with the new-physics scalar and pseudoscalar interactions are
\bea
H^{SP}_{1/2,0}&=&g_S \sqrt{Q_+}-g_P \sqrt{Q_-}, \nonumber\\
H^{SP}_{-1/2,0} &=& g_S\sqrt{Q_+}+g_P \sqrt{Q_-}.				
\eea
The parity-related amplitudes are
\bea
H^{S}_{\lambda_{c},\lambda_{NP}} & = & H^{S}_{-\lambda_{c},-\lambda_{NP}},\nonumber\\
H^{P}_{\lambda_{c},\lambda_{NP}} & = & -H^{P}_{-\lambda_{c},-\lambda_{NP}}.
\eea
For the vector and axial-vector helicity amplitudes, we find
\begin{align}
H^{VA}_{1/2,0}&= (1+g_L+g_R)\frac{\sqrt{Q_-}}{\sqrt{q^2}}(m_{b}+m_{c})-(1+g_L-g_R)\frac{\sqrt{Q_+}}{\sqrt{q^2}}(m_{b}-m_{c}), \nonumber\\
H^{VA}_{1/2,+1}&= -(1+g_L+g_R)\sqrt{2Q_-} + (1+g_L-g_R)\sqrt{2Q_+}, \nonumber\\	
H^{VA}_{1/2,t}&= (1+g_L+g_R)\frac{\sqrt{Q_+}}{\sqrt{q^2}}(m_{b}-m_{c})-(1+g_L-g_R)\frac{\sqrt{Q_-}}{\sqrt{q^2}}(m_{b}+m_{c}), \nonumber\\
H^{VA}_{-1/2,0}&= (1+g_L+g_R)\frac{\sqrt{Q_-}}{\sqrt{q^2}}(m_{b}+m_{c})+(1+g_L-g_R)\frac{\sqrt{Q_+}}{\sqrt{q^2}}(m_{b}-m_{c}), \nonumber\\	
H^{VA}_{-1/2,-1}&= -(1+g_L+g_R)\sqrt{2Q_-} -(1+g_L-g_R)\sqrt{2Q_+}, \nonumber\\		
H^{VA}_{-1/2,t}&= (1+g_L+g_R)\frac{\sqrt{Q_+}}{\sqrt{q^2}}(m_{b}-m_{c})+(1+g_L-g_R)\frac{\sqrt{Q_-}}{\sqrt{q^2}}(m_{b}+m_{c}).			
\end{align}
We also have the relations
\bea
H_{\lambda_{c},\lambda_{w}}^V&=&H_{-\lambda_{c},-\lambda_{w}}^V,\nonumber\\
H_{\lambda_{c},\lambda_{w}}^A&=&-H_{-\lambda_{c},-\lambda_{w}}^A.
\eea
The tensor helicity amplitudes are
\begin{align}
H^{(T)-1/2}_{-1/2,t,0}&= -g_T\big[-\sqrt{Q_-}+\sqrt{Q_+}\big], \nonumber\\[10pt]
H^{(T)+1/2}_{+1/2,t,0}&= g_T\big[\sqrt{Q_-}+\sqrt{Q_+}\big],\nonumber\\[10pt] 
H^{(T)-1/2}_{+1/2,t,+1}&= -g_T\frac{\sqrt{2}}{\sqrt{q^2}}\big[(m_{b}+m_{c})\sqrt{Q_-}+(m_{b}-m_{c})\sqrt{Q_+}\big],\nonumber\\[10pt] 
H^{(T)+1/2}_{-1/2,t,-1}&= -g_T\frac{\sqrt{2}}{\sqrt{q^2}}\big[(m_{b}+m_{c})\sqrt{Q_-}-(m_{b}-m_{c})\sqrt{Q_+}\big],\nonumber\\[10pt]
H^{(T)-1/2}_{+1/2,0,+1}&= -g_T\frac{\sqrt{2}}{\sqrt{q^2}}\big[(m_{b}+m_{c})\sqrt{Q_-}+(m_{b}-m_{c})\sqrt{Q_+}\big],\nonumber\\[10pt] 
H^{(T)+1/2}_{-1/2,0,-1}&= g_T\frac{\sqrt{2}}{\sqrt{q^2}}\big[(m_{b}+m_{c})\sqrt{Q_-}-(m_{b}-m_{c})\sqrt{Q_+}\big],\nonumber\\[10pt] 
H^{(T)+1/2}_{+1/2,+1,-1}&= -g_T\big[\sqrt{Q_-}+\sqrt{Q_+}\big],\nonumber\\[10pt] 
H^{(T)-1/2}_{-1/2,+1,-1}&= -g_T\big[\sqrt{Q_-}-\sqrt{Q_+}\big]. 
\end{align}
The other nonvanishing helicity amplitudes of tensor type are related to the above by
\bea
H^{(T)\lambda_{b}}_{\lambda_{c},\lambda,\lambda^\prime}=-H^{(T)\lambda_{b}}_{\lambda_{c},\lambda^\prime,\lambda}.
\eea

\subsection{Leptonic helicity amplitudes}

In the following, we define
\bea
v=\sqrt{1-\frac{m_\tau^2}{q^2}}.
\eea
The scalar and pseudoscalar leptonic helicity amplitudes are
\bea
L^{+1/2}=& 2\sqrt{q^2} v, \nonumber\\
L^{-1/2}=& 0,
\eea
while the vector and axial-vector amplitudes are
\bea
L^{+1/2}_{\pm1}&=&\pm\sqrt{2}m_{\tau} v\; \sin(\theta_\tau), \nonumber\\
L^{+1/2}_{0}&=&-2m_\tau v\; \cos{(\theta_\tau)}, \nonumber\\
L^{+1/2}_{t}&=&2m_\tau v, \nonumber\\
L^{-1/2}_{\pm1}&=&\sqrt{2 q^2}v\; (1\pm \cos(\theta_\tau)), \nonumber\\
L^{-1/2}_{0}&=&2\sqrt{q^2}v \; \sin{(\theta_\tau)}, \nonumber\\
L^{-1/2}_{t}&=&0,
\eea
and the tensor amplitudes are
\bea
L^{+1/2}_{0,\pm1}&=&-\sqrt{2 q^2}v\; \sin(\theta_\tau), \nonumber\\
L^{+1/2}_{\pm1,t}&=&\mp\sqrt{2 q^2}v\; \sin(\theta_\tau), \nonumber\\
L^{+1/2}_{t,0}&=&L^{+1/2}_{+1,-1}=-2\sqrt{q^2}v\; \cos(\theta_\tau), \nonumber\\
L^{-1/2}_{0,\pm1}&=&\mp\sqrt{2}m_\tau v\; (1\pm \cos(\theta_\tau)), \nonumber\\
L^{-1/2}_{\pm1,t}&=&-\sqrt{2}m_\tau v\; (1\pm \cos(\theta_\tau)), \nonumber\\
L^{-1/2}_{t,0}&=&L^{-1/2}_{+1,-1}=2m_\tau v\; \sin(\theta_\tau) .
\eea
Here we have the relation
\begin{equation}
 L^{\lambda_\tau}_{\lambda,\lambda^\prime}=-L^{\lambda_\tau}_{\lambda^\prime,\lambda}.
\end{equation}
%


\section{Four-body decay kinematics}
\label{appendix:kinematics}
In this appendix we derive the expression for the lepton's energy in the $b$ quark rest frame $E_\ell$, in terms of the scattering angle in the dilepton's rest frame $\theta_\ell$.
Consider the four-body decay
\begin{equation}
 b(p_{b})\rightarrow\ell^{-}(p_{\ell})+\bar{\nu}_{\ell}(p_{\bar{\nu}_\ell})+c(p_{c})+g(p_{g}),
\end{equation}
where $g$ is the real gluon. A four-body decay can be described in five invariants; here we present three of them which are relevant to our discussion. We have
\begin{align}
\label{eq:rsqrd}
r^2=& (p_c+p_g)^2=(p_b-p_\ell-p_\nu)^2,   \\
\label{eq:qsqrd}
q^2=& (p_\ell+p_\nu)^2=(p_b-p_g-p_c)^2,   \\
\label{eq:ssqrd}
s^2=& (p_b-p_\ell)^2=(p_g+p_c+p_\nu)^2.  
\end{align}

The expressions on the right-hand side above are written using 4-momentum conservation. By expanding Eq. (\ref{eq:ssqrd}) in the dilepton's rest frame we have
\begin{equation}
\label{eq:sxpand}
s^2=m_b^2+m_{\ell}^2-2 E^{\ell \nu}_b E^{\ell \nu}_\ell +2 P^{\ell \nu}_b P^{\ell \nu}_\ell cos(\theta_\ell), 
\end{equation}
where $E^{\ell \nu}_b$, $E^{\ell \nu}_\ell$, $P^{\ell \nu}_b$ and $P^{\ell \nu}_\ell$ refer to the energies and momenta of the $b$ quark and the massive lepton in the dilepton's rest frame. In order to find for these values in terms of invariants we expand Eq. (\ref{eq:rsqrd}), and using Eq. (\ref{eq:qsqrd}) we find
\begin{equation}
E^{\ell \nu}_b=\frac{m_b^2+q^2-r^2}{2\sqrt{q^2}}.
\end{equation}
One can also find
\begin{equation}
E^{\ell \nu}_\ell =\frac{m_{\ell}^2+q^2}{2\sqrt{q^2}}.
\end{equation}
Using the above expressions for energies we can easily find the corresponding momenta
\bea
P^{\ell \nu}_b &=& \frac{\sqrt{\lambda(m_b^2,q^2,r^2)}}{2\sqrt{q^2}},\\
P^{\ell \nu}_\ell &=& \frac{q^2-m_{\ell}^2}{2\sqrt{q^2}}, 
\eea
where $\lambda$ is defined as $\lambda(a,b,c)=a^2+b^2+c^2-2ab-2ac-2bc$.
Finally by expanding Eq. (\ref{eq:ssqrd}) again, but this time in the $b$ quark's rest frame, and using Eq. (\ref{eq:sxpand}) we find the expression for the lepton's energy as
\bea
E_\ell = \frac{1}{4 m_b q^2}\big[(m_b^2+q^2-r^2)(m_{\ell}^2+q^2)-(q^2-m_{\ell}^2)\sqrt{\lambda(m_b^2,q^2,r^2)}cos(\theta_\ell)\big].
\eea
In the case of three-body decay $b(p_{b})\rightarrow\ell^{-}(p_{\ell})+\bar{\nu}_{\ell}(p_{\bar{\nu}_\ell})+c(p_{c})$, $r^2$ reduces to $m_c^2$.

\section{Results for various observables}
\label{appendix:distributions}
For the twofold distribution $\frac{d\Gamma}{dq^2 dE_\ell}$, one finds from Eqs. (\ref{eq:rate}) and (\ref{Etau-relation})

\begin{eqnarray}
\label{twofold}
\frac{d \Gamma}{dq^2 dE_\ell} &=& \frac{G_F^2 |V_{cb}|^2 q^2 (1-m_\ell^2 /q^2)}{256 m_b^2 \pi^3}\bigg[ C_1^{VA} +\frac{m_\ell^2}{q^2} C_2^{VA} + C_3^{SP} \nonumber \\
 			   			   & &+ C_4^{T} + \frac{m_\ell^2}{q^2} C_5^{T} +\frac{4 m_\ell}{\sqrt{q^2}} C_6^{VA-SP} + \frac{8 m_\ell}{\sqrt{q^2}} C_7^{VA-T} + C_8^{SP-T} \bigg]
\end{eqnarray}
where the $C$ terms are

\begin{align}
C_1^{VA} =&~(1 + cos\theta)^2 | H^{VA}_{1/2,1}|^2 + (1 - cos\theta)^2 | H^{VA}_{-1/2,-1}|^2 + 2 sin\theta^2 | H^{VA}_{-1/2,0}|^2 + 2 sin\theta^2 | H^{VA}_{1/2,0}|^2, \nonumber \\[8pt]
C_2^{VA} =&~sin\theta^2 |H^{VA}_{1/2,1}|^2 + sin\theta^2 |H^{VA}_{-1/2,-1}|^2 + 2 | H^{VA}_{1/2,t} + cos\theta H^{VA}_{1/2,0}|^2 + 2 | H^{VA}_{-1/2,t} + cos\theta H^{VA}_{-1/2,0}|^2, \nonumber \\[8pt]
C_3^{SP} =&~2|H^{SP}_{1/2,0}|^2 + 2 |H^{SP}_{-1/2,0}|^2, \nonumber \\[8pt]
C_4^{T}  =&~8 cos\theta^2 | H^{(T)1/2}_{1/2,0,t} + H^{(T)1/2}_{1/2,1,-1}|^2 + 4 sin\theta^2 | H^{(T)1/2}_{-1/2,-1,t} + H^{(T)1/2}_{-1/2,0,-1} |^2  \nonumber \\[6pt]
		  &+ 4 sin\theta^2 |  H^{(T)-1/2}_{1/2,t,1} + H^{(T)-1/2}_{1/2,0,1} |^2 +8 cos\theta^2 | H^{(T)-1/2}_{-1/2,0,t} + H^{(T)-1/2}_{-1/2,1,-1}|^2,  \nonumber \\[8pt]
C_5^{T}  =&~8 sin\theta^2 | H^{(T)1/2}_{1/2,0,t} + H^{(T)1/2}_{1/2,1,-1}|^2 +4(1 - cos\theta)^2 | H^{(T)1/2}_{-1/2,-1,t} + H^{(T)1/2}_{-1/2,0,-1}|^2 \nonumber \\[6pt]
		  &+ 4(1 + cos\theta)^2 | H^{(T)-1/2}_{1/2,0,1} + H^{(T)-1/2}_{1/2,t,1}|^2 + 8 sin\theta^2 | H^{(T)-1/2}_{-1/2,0,t} + H^{(T)-1/2}_{-1/2,1,-1}|^2,  \nonumber \\[8pt]
C_6^{VA-SP} =&~Re[(cos\theta H^{VA}_{1/2,0}+H^{VA}_{1/2,t})H^{SP *}_{1/2,0}] + Re[(cos\theta H^{VA}_{-1/2,0}+H^{VA}_{-1/2,t})H^{SP *}_{-1/2,0}], \nonumber \\[8pt]
C_7^{VA-T}  =&~(1+cos\theta)Re[(H^{(T)-1/2}_{1/2,0,1}+H^{(T)-1/2}_{1/2,t,1})H^{VA *}_{1/2,1}] -(1-cos\theta) Re[( H^{(T)1/2}_{-1/2,-1,t} + H^{(T)1/2}_{-1/2,0,-1})H^{VA *}_{-1/2,-1}] \nonumber \\[6pt]
			 &-Re[(H^{(T)1/2}_{1/2,0,t} + H^{(T)1/2}_{1/2,1,-1})(H^{VA *}_{1/2,0}+cos\theta H^{VA *}_{1/2,t})] \nonumber \\[6pt]
			 &-Re[( H^{(T)-1/2}_{-1/2,0,t} + H^{(T)-1/2}_{-1/2,1,-1})( cos\theta H^{VA *}_{-1/2,t} + H^{VA *}_{-1/2,0})], \nonumber \\[8pt]
C_8^{SP-T}  =&~-8 cos\theta Re[H^{SP *}_{1/2,0}(H^{(T)1/2}_{1/2,0,t}+H^{(T)1/2}_{1/2,1,-1})] -8 cos\theta Re[H^{SP *}_{-1/2,0}(H^{(T)-1/2}_{-1/2,0,t}+H^{(T)-1/2}_{-1/2,1,-1})], 
\end{align}
with
\begin{equation}
cos\theta = \frac{(m_b^2-m_c^2+q^2)(q^2+m_\ell^2)-(4m_b q^2 E_\ell)}{\sqrt{Q_+Q_-}(q^2-m_\ell^2)}.
\end{equation}
From relation (\ref{twofold}), one can conveniently find the distribution for $q^2$ or $E_\ell$. Nonperturbative corrections to these distributions (for SM) are presented elsewhere (see \cite{Falk:1994gw}, \cite{Ligeti:2014kia} and \cite{Balk:1993sz}) and we do not repeat them here.\\
The forward-backward asymmetry can be written as the sum of tree level $A_{FB}^0$ and nonperturbative $A_{FB}^{\mathcal{O}(1/m_b^2)}$ terms,
\begin{equation}
A_{FB} = A_{FB}^0 + A_{FB}^{\mathcal{O}(1/m_b^2)},
\end{equation}
with
\begin{eqnarray}
A_{FB}^0 &=& (\frac{d\Gamma}{dq^2})^{-1}~\frac{G_F^2 |V_{cb}|^2}{512 \pi^3}\frac{q^2 \sqrt{Q_+Q_-}}{m_b^3}\Big(1-\frac{m_{\ell}^2}{q^2}\Big)^2 \Bigg[ B_1^{VA}+\frac{2m_{\ell}^2}{q^2}B_2^{VA}+\frac{4m_{\ell}^2}{q^2}B_3^T+\nonumber\\[5pt]
&&\frac{2m_\ell}{\sqrt{q^2}} B_4^{VA-SP}+\frac{4m_\ell}{\sqrt{q^2}} B_5^{VA-T}+4B_6^{SP-T} \Bigg],
\end{eqnarray}
where
\begin{align}
B_1^{VA}=&~|H^{VA}_{1/2,1}|^2-|H^{VA}_{-1/2,-1}|^2,\nonumber\\[10pt]
B_2^{VA}=&~\mathrm{Re}[H^{VA*}_{1/2,t}H^{VA}_{1/2,0}+H^{VA*}_{-1/2,t}H^{VA}_{-1/2,0}],\nonumber\\[10pt]
B_3^{T}=&~|H^{(T)-1/2}_{1/2,0,1}+H^{(T)-1/2}_{1/2,t,1}|^2-|H^{(T)1/2}_{-1/2,-1,0}+H^{(T)1/2}_{-1/2,t,-1}|^2,\nonumber\\[10pt]
B_4^{VA-SP}=&~\mathrm{Re}[H^{SP*}_{1/2,0}H^{VA}_{1/2,0}+H^{SP*}_{-1/2,0}H^{VA}_{-1/2,0}],\nonumber\\[10pt]
B_5^{VA-T}=&~\mathrm{Re}[H^{VA*}_{1/2,t} (H^{(T)1/2}_{1/2,-1,1}+H^{(T)1/2}_{1/2,t,0})]+\mathrm{Re}[H^{VA*}_{1/2,1} (H^{(T)-1/2}_{1/2,0,1}+H^{(T)-1/2}_{1/2,t,1})]\nonumber\\[10pt]
&+\mathrm{Re}[H^{VA*}_{-1/2,t} (H^{(T)-1/2}_{-1/2,-1,1}+H^{(T)-1/2}_{-1/2,t,0})]-\mathrm{Re}[H^{VA*}_{-1/2,-1} (H^{(T)1/2}_{-1/2,-1,0}+H^{(T)1/2}_{-1/2,t,-1})],\nonumber\\[10pt]
B_6^{SP-T}=&~\mathrm{Re}[H^{SP*}_{1/2,0}(H^{(T)1/2}_{1/2,-1,1}+H^{(T)1/2}_{1/2,t,0})]+\mathrm{Re}[H^{SP*}_{-1/2,0}(H^{(T)-1/2}_{-1/2,-1,1}+H^{(T)-1/2}_{-1/2,t,0})].
\end{align}

Also, the $\mathcal{O}(1/m_b^2)$ correction is
\begin{eqnarray}
A_{FB}^{\mathcal{O}(1/m_b^2)} &=& \big( \frac{d\Gamma}{dq^2} \big)^{-1}\frac{G_F^2 | V_{cb}|^2(1-m_\ell^2 /q^2)^2}{384\pi^3 m_b^5 q^2}\big\{ \lambda_1[(m_\ell ^2 m_b^2 -m_\ell ^2 m_c^2 -(q^2)^2 )(3(m_b^2-m_c^2)^2 \nonumber \\[6pt]  
							  & & +q^2(2m_b^2-6m_c^2+3q^2))] + \lambda_2[9m_b^6 m_\ell ^2 -45(m_c^2-q^2)^2(m_c^2 m_\ell ^2 +(q^2)^2) \nonumber \\[6pt]
	         				  & & +m_b^4(-63m_c^2 m_\ell ^2+3q^2(2m_\ell ^2+9q^2)) +3m_b^2(33m_c^4 m_\ell ^2+2m_c^2 q^2(-8m_\ell ^2+3q^2) \nonumber \\[6pt]
	         				  & & +(q^2)^2(3m_\ell ^2+14 q^2))]\big\}.
\end{eqnarray}

\pagebreak

\end{document}